\documentclass[12pt,fleqn]{article}
\usepackage{amsmath,amssymb,amsthm,tikz,indentfirst,graphicx,mathrsfs,enumerate,caption,float,authblk,subcaption}
\usepackage[square,comma,numbers,sort&compress]{natbib}
\usepackage[colorlinks]{hyperref}
 \theoremstyle{plain}
\newtheorem{theorem}{Theorem}[section]
\newtheorem{corollary}[theorem]{Corollary}
\newtheorem{lemma}[theorem]{Lemma}
\newtheorem{assume}[theorem]{Assumption}
\newtheorem{prop}[theorem]{Proposition}
 \theoremstyle{remark}
\newtheorem{remark}[theorem]{Remark}
 \theoremstyle{definition}
 \newtheorem{definition}[theorem]{Definition}
\newtheorem{problem}[theorem]{Riemann--Hilbert Problem}
\usepackage[left=2.5cm,right=2.5cm,bottom=2.8cm,top=2.8cm]{geometry}

\newcommand{\res}{\operatorname{Res}\ }
\newcommand{\im}{\mathrm{i}}

\numberwithin{equation}{section}
\allowdisplaybreaks[2]
\begin{document}
\title{Multiple higher-order poles solutions in spinor  Bose-Einstein condensates}

\author[1]{Huan Liu}
\author[2]{Jing Shen}
\author[1]{Xianguo Geng\thanks{Corresponding author.
E-mail address: xggeng@zzu.edu.cn}}
\affil[1]{School of Mathematics and Statistics, Zhengzhou University, Zhengzhou, Henan 450001, People's  Republic  of China}
\affil[2]{School  of  Sciences,  Henan  University  of  Technology,  Zhengzhou, Henan  450001,  People's  Republic  of China}

\renewcommand*{\Affilfont}{\small\it}
\renewcommand{\Authands}{, }
\date{}

  \maketitle
  \begin{abstract}
 In this study, we explore multiple higher-order pole solutions in spinor Bose--Einstein condensates. These solutions are associated with different pairs of higher-order poles of the transmission coefficient in the inverse scattering transform, and they represent solutions of the spin-1 Gross--Pitaevskii equation. We introduce a direct scattering map that maps initial data to scattering data, which includes discrete spectrums, reflection coefficients, and a polynomial that replaces normalization constants. To analyze symmetries and discrete spectrums in the direct problem, we introduce a generalized cross product in 4-dimensional vector space. Additionally, we characterize the inverse problem in terms of a $4\times 4$ matrix Riemann--Hilbert problem that is subject to residue conditions at these higher-order poles. In the reflectionless scenario, the Riemann--Hilbert problem can be converted  into a linear algebraic system. The resulting algebraic system has a unique solution, which allows us to display multiple higher-order poles  solutions.

 \end{abstract}


\newpage
\section{Introduction}
The experimental observation of Bose-Einstein condensates (BECs) \cite{AEMWC1995,DMADDKK1995} has led to extensive study of their physical properties. In single-component BECs, the Gross-Pitaevskii (GP) equation, also known as the nonlinear Schrödinger (NLS) equation in one dimension, is the relevant dynamic model. Multi-component BECs, which exhibit rich and diverse dynamics, have received significant attention since their experimental realization \cite{MMRRI2002,MBGCW1997,STIR2010}. As one of the completely integrable 3-component BEC models, the spin-1 GP equation was proposed to describe the soliton dynamics in spinor BECs \cite{IMW2004} and has been extensively studied. N-soliton solutions \cite{IMW2004,POVC2018,RRR2020,ZTQY2022} and rogue wave solutions \cite{LPB2018,QM2012} have been obtained using inverse scattering transform (IST) and Darboux transformation methods. The Riemann-Hilbert (RH) method has been used to investigate initial-boundary problems on the half-line and finite interval \cite{Yan2017,Yan2019}, IST with vanishing boundary conditions \cite{IUW2007,PDLH2018}, and long-time asymptotics without solitons \cite{GWC2021}.

On the other hand, the poles of the transmission coefficient in IST theory provide the mechanism that gives rise to soliton solutions of integrable equations. In fact, $N$ conjugate pairs of simple poles result in an $N$-soliton solution. Therefore, it is natural to investigate the possible reflectionless solutions of integrable equations with a transmission coefficient consisting of $N$ pairs of higher-order poles. These solutions are referred to as multiple higher-order poles solutions. Several integrable equations associated with $2\times2$ matrix spectral problems have been studied, such as the NLS equation \cite{Aktosun2007,Olmedilla1987,Schiebold2017,ZS1972,ZhangHe2020}, the modified Korteweg--de Vries equation \cite{Wadati1982,ZhangTao2020}, the sine-Gordon equation \cite{Wadati1984}, the $N$-wave system \cite{ShYang2003,SYang2003}, the derivative NLS equation \cite{ZY2020}, the modified short-pulse system \cite{LL2023,LQL2022}, discrete integrable equations \cite{CF2022,CFH2023}, and more. However, there has been little literature on higher-order poles solutions of integrable equations associated with $4\times 4$ matrix spectral problems.

In this work, we will formulate a RH problem equipped with several residue conditions at $N$ pairs of multiple poles  and  provide an algebraic characterization of  multiple higher-order poles  solutions for the spin-1 GP equation 
\begin{equation}\label{Eq:SGPE}
  \im \mathbf{Q}_t+\mathbf{Q}_{xx}+2\mathbf{Q}\mathbf{Q}^\dag\mathbf{Q}=\mathbf{0},\quad \mathbf{Q}=\begin{pmatrix}
    q_1&q_0\\
    q_0&q_{-1}
  \end{pmatrix},
\end{equation}
associated with a $4\times 4$ matrix spectral problem. Our findings are comprised of two key components. Firstly, we establish the dependence relations of Jost eigenfunctions at the discrete spectrum by introducing the generalized cross product in higher-dimensional Euclidean space. The critical aspect of this element is deriving the symmetries among these relations, as demonstrated in Corollary\,\ref{cormu} and its corresponding proof. Using the Hermite interpolation formula, we have unified the residue conditions at all multiple poles, as shown in Proposition\,\ref{prop:res}.
Secondly, under the assumption of reflectionlessness, we construct a concise linear algebraic system, and provide a rigorous proof that the resulting algebraic system possesses a unique solution by transforming the coefficient matrix into a Hermitian positive definite matrix in Section\,\ref{sect:regular}. Consequently, multiple higher-order poles solutions can be reconstructed by solving this linear system, with opportunities for obtaining more explicit solutions through appropriate parameter selection.

 The rest of this paper is organized as follows: In Section\,\ref{dp}, we establish a mapping from the initial data to the scattering data and analyze the discrete spectrum associated with multiple zeros. In Section\,\ref{inverse}, we establish a mapping from the scattering data to a $4\times $ matrix RH problem with $N$ pairs of multiple poles and characterize the solution to the spin-1 GP equation\,\eqref{Eq:SGPE} in terms of the solution to the RH problem. Under the assumption of reflectionlessness, multiple higher-order pole solutions can be obtained by solving a linear algebraic system which has been proved to solved uniquely in Section\,\ref{sect:regular}, and we display the density structures of some multipole solutions in Section\,\ref{sect:explicit}.
Throughout the paper, we use the following notations: The complex conjugate of a complex number $\lambda$ is denoted by $\bar\lambda$. For a complex-valued matrix $\mathbf{A}$, $\bar{\mathbf{A}}$  denotes the element-wise complex conjugate, $\mathbf{A}^{\mathrm{T}}$  denotes the transpose, and $\mathbf{A}^{\dag}$ denotes the conjugate transpose. A $4\times 4$ matrix $\mathbf{A}$  is represented as four blocks:
 \[\mathbf{A}=\begin{pmatrix}
                \mathbf{A}_{11} & \mathbf{A}_{12} & \mathbf{A}_{13}&  \mathbf{A}_{14}\\
                \mathbf{A}_{21} & \mathbf{A}_{22} & \mathbf{A}_{23}& \mathbf{A}_{24} \\
                \mathbf{A}_{31} & \mathbf{A}_{32} & \mathbf{A}_{33}& \mathbf{A}_{34}\\
                 \mathbf{A}_{41} & \mathbf{A}_{42} & \mathbf{A}_{43}& \mathbf{A}_{44}
              \end{pmatrix}=(\mathbf{A}_1,\mathbf{A}_2,\mathbf{A}_3,\mathbf{A}_4)=(\mathbf{A}_{\mathrm{L}},\mathbf{A}_{\mathrm{R}})=\begin{pmatrix}
  \mathbf{A}_{\mathrm{UL}}&\mathbf{A}_{\mathrm{UR}}\\\mathbf{A}_{\mathrm{DL}}&\mathbf{A}_{\mathrm{DR}}
\end{pmatrix},\]
 where  $\mathbf{A}_{ij}$ represents the $(i,j)$-entry, $\mathbf{A}_{ j}$ represents the $j$-th column, $\mathbf{A}_{\mathrm{L}}$ represents the first two columns, $\mathbf{A}_{\mathrm{R}}$ represents  the  last two columns, $\mathbf{A}_{\mathrm{UL}},\mathbf{A}_{\mathrm{UR}},\mathbf{A}_{\mathrm{DL}},\mathbf{A}_{\mathrm{DR}}$ are $2\times 2$ matrices. The notation $\mathbf{A}(\lambda),~\lambda\in(D_1,D_2)$, means that $ \mathbf{A}_{\mathrm{L}}$ and $\mathbf{A}_{\mathrm{R}}$  holds for $\lambda\in D_1$, $D_2$, respectively. Furthermore, $\sigma_1=\begin{pmatrix}
  0&1\\
  1&0
\end{pmatrix}$,
$\sigma_2=\begin{pmatrix}
  0&-\mathrm{i}\\
  \mathrm{i}&0
\end{pmatrix}$, $\sigma_3=\begin{pmatrix}
  1&0\\
  0&-1
\end{pmatrix}$ represent three type of Pauli matrices,  $\mathbf{I}_n$ represents the $n\times n$ identity matrix, and
$\mathbb{C}^\pm=\{k\in\mathbb{C}:\operatorname{Im} k\gtrless0\}$. For a (vector-valued) function $\mathbf{f}(x,t;\lambda)$,  $\mathbf{f}^{(n)}(x,t;\lambda)=\partial_{\lambda}^n\mathbf{f}(x,t;\lambda)$, $\mathbf{f}^{(n)}(x,t;\lambda_0)=\partial_{\lambda}^n\mathbf{f}(x,t;\lambda)|_{\lambda=\lambda_0}$.

\section{Direct problem}\label{dp}
\subsection{Jost solutions and scattering matrix}
The spin-1 GP equation \eqref{Eq:SGPE} admits  the Lax pair:
\begin{subequations}\label{eq:Laxpair}
  \begin{align}
    &\psi_x=(-\mathrm{i}\lambda\sigma+\mathbf{U})\psi,\label{subeq:lpx}\\
   & \psi_t=(-2\mathrm{i}\lambda^2\sigma+\tilde{\mathbf{U}})\psi,
  \end{align}
\end{subequations}
where $\psi(x,t;\lambda)$ is a  matrix-valued function of $x,t$ and the spectral parameter $\lambda\in\mathbb C$,
\[\sigma=\begin{pmatrix}
  \mathbf{I}_2&\mathbf{0}\\\mathbf{0}&-\mathbf{I}_2
\end{pmatrix},\quad \mathbf{U}=\begin{pmatrix}
  \mathbf{0}&\mathbf{Q}\\-\mathbf{Q}^{\dag}&\mathbf{0}
\end{pmatrix},\quad \tilde{\mathbf{U}}=2\lambda\mathbf{U}+\mathrm{i}\sigma(\mathbf{U}_x-\mathbf{U}^2).\]
Indeed, $
  \sigma=\sigma_3\otimes\mathbf{I}_2$,
 where $``\otimes "$ represents   Kronecker product.

Assuming that  $\mathbf{Q}(x,t)$  approaches zero as $x$ becomes large, and that its $x$-derivative also vanishes at infinity, we seek two Jost solutions $\psi_\pm(x,t;\lambda)$ of Eq.\,\eqref{eq:Laxpair}. These solutions satisfy boundary conditions
\begin{equation}\label{eq:Psiasy}
  \psi_\pm(x,t;\lambda)= \mathrm{e}^{-\mathrm{i}\theta(x,t;\lambda)\sigma}+o(1), \quad x\rightarrow \pm\infty,
\end{equation}
where  $\theta(x,t;\lambda)=\lambda x+2\lambda^2 t$. As  $\mathbf{U}(x,t)$ is traceless,  we can use Abel's Theorem to derive  $\partial_x\det( \psi_\pm(x,t;\lambda))=0$. This gives us
 \begin{equation}\label{psidet}
   \det(\psi_\pm(x,t;\lambda))=1,\quad \lambda\in\mathbb{R}.
 \end{equation}
 Since both $\psi_+(x,t;\lambda)$ and $\psi_-(x,t;\lambda)$ are fundamental solutions of the Lax pair \eqref{eq:Laxpair},   a $4\times4$ scattering  matrix $\mathbf{S}(\lambda)$ exists that is independent of $x$ and $t$ and satisfies
\begin{equation}\label{eq:Jostjump}
  \psi_-(x,t;\lambda)=\psi_+(x,t;\lambda)\mathbf{S}(\lambda),\quad \lambda\in\mathbb{R}.
\end{equation}
 Combining Eq.\,\eqref{eq:Jostjump} with the condition in  Eq.\,\eqref{psidet} yields
\begin{equation}
 \det(\mathbf{S}(\lambda))=1,\quad \lambda\in\mathbb{R}.
\end{equation}

Let
\begin{equation}\label{mJost}
  \mu_\pm(x,t;\lambda)=\psi_\pm(x,t;\lambda)\mathrm{e}^{\mathrm{i}\theta(x,t;\lambda)\sigma},
\end{equation}
then
\begin{equation}\label{mux}
   \partial_x\mu_\pm=[-\mathrm{i}\lambda\sigma, \mu_\pm]+\mathbf{U}\mu_\pm,\quad \lim_{x\rightarrow \pm\infty }\mu_\pm=\mathbf{I}_4,
\end{equation}
where $[\cdot,\cdot]$ represents a  commutator, and the $(x,t;\lambda)$-dependence is omitted for brevity.
Eq.\,\eqref{mux} is  equivalent to
\begin{equation}\label{eq:mupmin}
  \mu_\pm(x,t;\lambda)=\mathbf{I}_4+\int_{\pm\infty}^x\mathrm{e}^{\mathrm{i}\lambda(x'-x)\hat{\sigma}}(\mathbf{U}\mu_\pm)(x',t;\lambda)\mathrm{d}x',
  \end{equation}
where $\hat{\sigma}\mathbf{X}=[\sigma,\mathbf{X}]$, consequently, $\mathrm{e}^{\hat{\sigma}}\mathbf{X}=\mathrm{e}^{\sigma}\mathbf{X}\mathrm{e}^{-\sigma}$.

By using standard Volterra theory, we can make  the following statement.

\begin{theorem}\label{Th-ana}
   Assuming that $ \mathbf{Q}(\cdot ,t)\in L^{1}(\mathbb{R})$ for fixed $t$, the modified eigenfunctions $\mu_\pm(x,t;\lambda)$ in Eq.\,\eqref{eq:mupmin} are well defined for $\lambda\in\mathbb{R}$, $\mu_{-\mathrm{L}}$ and $\mu_{+\mathrm{R}}$ can be analytically continued onto the upper half-plane $\mathbb{C}^+$, $\mu_{+\mathrm{L}}$ and $\mu_{-\mathrm{R}}$ can be analytically continued onto the lower half-plane $\mathbb{C}^-$. For all $\lambda$ within the interior of their corresponding domains of analyticity,  $\mu_{\pm}(x,t;\lambda)$ are bounded for $x\in\mathbb{R}$. Furthermore, the same analyticity properties hold for  $\psi_\pm(x,t;\lambda)$.
\end{theorem}

\subsection{Symmetry and asymptotic behavior}

We have various symmetries for the Lax pair  \eqref{eq:Laxpair} given as
\[\mathbf{U}^{\dag}=-\mathbf{U},\quad \tau \mathbf{U}\tau=\bar{\mathbf{U}},\quad \tau=\begin{pmatrix}
 \mathbf{0}&-\mathrm{i}\mathbf{I}_2\\
 \mathrm{i}\mathbf{I}_2&\mathbf{0}
\end{pmatrix}=\sigma_2\otimes\mathbf{I}_2.\]
These lead to certain relationships between   $\psi_{\pm}(x,t;\lambda)$,  $ [\psi_{\pm}^{\dag}(x,t;\bar\lambda)]^{-1}$ and $\tau\bar\psi_{\pm}(x,t;\bar\lambda)\tau$ that satisfy the same differential equation \eqref{eq:Laxpair}.
Using the asymptotic conditions \eqref{eq:Psiasy} for $\psi_\pm(x,t;\lambda)$, we can show that
  \begin{equation}\label{eq:Jostsym1}
   \psi_{\pm}^{\dag}(x,t;\bar\lambda)=\psi_{\pm}^{-1}(x,t;\lambda), \quad \psi_{\pm}(x,t;\lambda)=\tau\bar\psi_{\pm}(x,t;\bar\lambda)\tau,\quad \lambda\in\mathbb{R}.
  \end{equation}
 In addition, the following symmetric and conjugate properties hold for $\mathbf{S}(\lambda)$:
  \begin{equation}\label{eq:ssym}
 \mathbf{S}^{-1}(\lambda)=\mathbf{S}^\dag(\bar\lambda),\quad \mathbf{S}(\lambda)=\tau\bar{\mathbf{S}}(\bar\lambda)\tau, \quad \lambda\in\mathbb{R},
  \end{equation}
 i.e.,
\begin{equation}\label{eq:ssym1}
  \mathbf{S}_{\mathrm{UL}}(\lambda)=\bar{\mathbf{S}}_{\mathrm{DR}}(\bar\lambda),\quad \mathbf{S}_{\mathrm{DL}}(\lambda)=-\bar{\mathbf{S}}_{\mathrm{UR}}(\bar\lambda),
\end{equation}
which means that we can express $\mathbf{S}(\lambda)$ as
\begin{equation}\label{eq:sabr}
\mathbf{S}(\lambda)=\begin{pmatrix}
  \mathbf{a}(\lambda)&-\bar{\mathbf{b}}(\bar\lambda)\\ \mathbf{b}(\lambda)& \bar{\mathbf{a}}(\bar\lambda)
\end{pmatrix}.
\end{equation}
Furthermore, it can be inferred from Eq.\,\eqref{eq:ssym} that for $\lambda\in\mathbb{R}$, we have
\begin{equation}
\begin{aligned}
 &\mathbf{a}^\dag(\bar\lambda)\mathbf{a}(\lambda)+ \mathbf{b}^\dag(\bar\lambda)\mathbf{b}(\lambda)=\mathbf{I}_2, \quad&& \mathbf{a}^{\mathrm{T}}(\lambda)\mathbf{b}(\lambda)=\mathbf{b}^{\mathrm{T}}(\lambda)\mathbf{a}(\lambda),\\
  &\mathbf{a}(\lambda)\mathbf{a}^\dag(\bar\lambda)+ \bar{\mathbf{b}}(\bar\lambda)\mathbf{b}^{\mathrm{T}}(\lambda)=\mathbf{I}_2, \quad && \mathbf{a}(\lambda)\mathbf{b}^\dag(\bar\lambda)=\bar{\mathbf{b}}(\bar\lambda)\mathbf{a}^{\mathrm{T}}(\lambda).
\end{aligned}
\end{equation}
Using Eq.\,\eqref{eq:Jostjump} and Eq.\,\eqref{eq:mupmin}, we note that $\mathbf{a}(\lambda)$ and $\mathbf{b}(\lambda)$ satisfy
\begin{equation}
  \mathbf{a}(\lambda)=\mathbf{I}_2+\int_{-\infty}^{+\infty}\mathbf{Q}(x,0)\mu_{-21}(x,0;\lambda)\mathrm{d}x,\quad
  \mathbf{b}(\lambda)=-\int_{-\infty}^{+\infty}\mathrm{e}^{-2\mathrm{i}\lambda x}\mathbf{Q}^\dag(x,0)\mu_{-11}(x,0;\lambda)\mathrm{d}x.
\end{equation}
Suppose that $ \mathbf{Q}(x,0)\in \mathrm{L}^{1}(\mathbb{R})$,   $\mathbf{a}(\lambda)$ and  $\mathbf{b}(\lambda)$  are well defined for $\lambda\in\mathbb{R}$, $\mathbf{a}(\lambda)$ can be analytically continued onto $\mathbb{C}^+$.
   Furthermore, it follows from Eqs.\,\eqref{eq:Jostjump}, \eqref{mJost}, \eqref{eq:Jostsym1} and \eqref{eq:sabr} that $\mathbf{a}(\lambda)$ and  $\mathbf{b}(\lambda)$ can be expressed in terms  of $\mu_\pm(x,t;\lambda)$ or $\psi_\pm(x,t;\lambda)$:
 \begin{subequations}\label{eq:sS}
 \begin{alignat}{2}
 &\mathbf{a}(\lambda) =\mu_{+\mathrm{L}}^\dag(\bar\lambda)\mu_{-\mathrm{L}}(\lambda)=\psi_{+\mathrm{L}}^\dag(\bar\lambda)\psi_{-\mathrm{L}}(\lambda),&&\  \lambda\in \mathbb{C}^+\cup \mathbb{R},\label{eq:sSa}\\
 &\det[\mathbf{a}(\lambda)]=\det[\mu_{-\mathrm{L}}(\lambda),\mu_{+\mathrm{R}}(\lambda)]=\det[\psi_{-\mathrm{L}}(\lambda),\psi_{+\mathrm{R}}(\lambda)],
\quad  &&\  \lambda\in \mathbb{C}^+\cup \mathbb{R},\\
 &\bar{\mathbf{a}}(\bar\lambda) =\mu_{+\mathrm{R}}^\dag(\bar\lambda)\mu_{-\mathrm{R}}(\lambda)=\psi_{+\mathrm{R}}^\dag(\bar\lambda)\psi_{-\mathrm{R}}(\lambda),&& \  \lambda\in \mathbb{C}^-\cup \mathbb{R},\\
 &
 \det[\bar{\mathbf{a}}(\bar\lambda)]=\det[\mu_{+\mathrm{L}}(\lambda),\mu_{-\mathrm{R}}(\lambda)]=\det[\psi_{+\mathrm{L}}(\lambda),\psi_{-\mathrm{R}}(\lambda)],
 && \  \lambda\in \mathbb{C}^-\cup \mathbb{R},\label{eq:sSb}\\
 &\mathbf{b}(\lambda)= \mathrm{e}^{-2\mathrm{i}\theta(\lambda)}\mu_{+\mathrm{R}}^\dag(\lambda)\mu_{-\mathrm{L}}(\lambda)=\psi_{+\mathrm{R}}^\dag(\lambda)\psi_{-\mathrm{L}}(\lambda),
  &&\  \lambda\in  \mathbb{R},\\
  &
 -\bar{\mathbf{b}}(\bar\lambda)=\mathrm{e}^{2\mathrm{i}\theta(\lambda)}\mu_{+\mathrm{L}}^\dag(\lambda)\mu_{-\mathrm{R}}(\lambda)
 =\psi_{+\mathrm{L}}^\dag(\lambda)\psi_{-\mathrm{R}}(\lambda),
  &&\  \lambda\in  \mathbb{R}.
  \end{alignat}
  \end{subequations}

Substituting the Wentzel--Kramers--Brillouin  expansions of  $\mu_{\pm}(x,t;\lambda)$ into Eq.\,\eqref{mux} and collecting the terms $O(\lambda^j)$, we can derive the following asymptotic expressions
\begin{equation}\label{asyeg}
  \begin{aligned}
  &(\mu_{+\mathrm{L}}(x,t;\lambda),\mu_{-\mathrm{R}}(x,t;\lambda))=\mathbf{I}_4+O(\lambda^{-1}),\quad &&\lambda\in\mathbb{C}_-\rightarrow \infty,\\
  &(\mu_{-\mathrm{L}}(x,t;\lambda),\mu_{+\mathrm{R}}(x,t;\lambda))=\mathbf{I}_4+O(\lambda^{-1}),\quad&& \lambda\in\mathbb{C}_+\rightarrow\infty.
  \end{aligned}
     \end{equation}
As a result,
\begin{equation}\label{aasy}
  \begin{aligned}
& \mathbf{a}(\lambda) =\mathbf{I}_2+O(\lambda^{-1}),\qquad&& \lambda\rightarrow \infty,\\
 &\mathbf{b}(\lambda) =O(\lambda^{-1}),\qquad&& \lambda\rightarrow \infty.
  \end{aligned}
 \end{equation}

We define the reflection coefficient
\begin{equation}\label{rho}
 \gamma(\lambda)=\mathbf{b}(\lambda)\mathbf{a}^{-1}(\lambda),\quad \lambda\in\mathbb{R},
  \end{equation}
The transmission coefficient is $\frac{1}{\det[\mathbf{a}(\lambda)]}$,
  and  we have $\gamma^\mathrm{T}(\lambda)=\gamma(\lambda)$, $\gamma(\lambda)=O(\lambda^{-1})$ as $\lambda\rightarrow\infty$.

\subsection{Discrete spectrum}
Combining Eq.\,\eqref{psidet} with Eq.\,\eqref{eq:Jostsym1} yields
\begin{equation}\label{eq:relacross}
  \psi_{\pm j_4}(x,t;\lambda)=\mathcal{G}[\bar{\psi}_{\pm j_1}(x,t;\bar\lambda), \bar{\psi}_{\pm j_2}(x,t;\bar\lambda),\bar{\psi}_{\pm j_3}(x,t;\bar\lambda)],\quad \lambda\in\mathbb{R},
\end{equation}
where   $(j_1,j_2,j_3,j_4)$  is an even permutation of $(1,2,3,4)$ and ``$\mathcal{G}[\cdot]$" represents the generalized cross product defined in Ref.\cite{LSG2023}, i.e., for all $\mathbf{u}_1,\mathbf{u}_2,\mathbf{u}_{3}\in\mathbb{C}^{4}$,
\begin{equation}
  \mathcal{G}[\mathbf{u}_1,\mathbf{u}_2,\mathbf{u}_3]=\sum_{j=1}^{4}\det(\mathbf{u}_1,\mathbf{u}_2,\mathbf{u}_3,\mathbf{e}_j)\mathbf{e}_j,\\
\end{equation}
where $\{\mathbf{e}_1,\mathbf{e}_2,\mathbf{e}_3,\mathbf{e}_4\}$ represents the standard basis for $\mathbb{R}^{4}$.
 By direct calculations, it is easy to verify the following relation between the adjugate matrix $(\cdot)^*$ and the generalized cross product $\mathcal{G}[\cdot]$,
 \begin{equation}\label{adjG}
   \mathbf{u}^*=(\mathbf{u}_1,\mathbf{u}_2,\mathbf{u}_3,\mathbf{u}_4)^*=\begin{pmatrix}
     -\mathcal{G}^{\mathrm{T}}[\mathbf{u}_2,\mathbf{u}_3,\mathbf{u}_4]\\
     \mathcal{G}^{\mathrm{T}}[\mathbf{u}_1,\mathbf{u}_3,\mathbf{u}_4]\\
    -\mathcal{G}^{\mathrm{T}}[\mathbf{u}_1,\mathbf{u}_2,\mathbf{u}_4]\\
     \mathcal{G}^{\mathrm{T}}[\mathbf{u}_1,\mathbf{u}_2,\mathbf{u}_3]
   \end{pmatrix}.
 \end{equation}
 \begin{lemma}\label{le:g}
   For  $\mathbf{u}_1,\mathbf{u}_2,\mathbf{u}_{3}\in\mathbb{C}^{4}$, $\mathcal{G}[\mathbf{u}_1,\mathbf{u}_2,\mathbf{u}_3]=\mathbf{0}$ if and only if  $\mathbf{u}_1,\mathbf{u}_2,\mathbf{u}_{3}$ are linearly dependent. Furthermore,  $\mathcal{G}[\cdot]$ is also multi-linear and totally antisymmetric.
 \end{lemma}
\begin{prop}\label{prop:psi}
Suppose that $\lambda_0$ is a zero of $\det[\mathbf{a}(\lambda)]$ with multiplicity $m+1$, then there exist $m+1$  complex-valued constant $2\times 2$  symmetric matrices $\mathbf{B}_0,\mathbf{B}_1,\ldots,\mathbf{B}_m$ with $\mathbf{B}_0$ not equaling the zero matrix, such that for each $n\in\{0,\ldots,m\}$,
 \begin{equation}\label{eq:psid}
  \frac{ [\psi_{-\mathrm{L}}(x,t;\lambda_0)\mathrm{adj}[\mathbf{a}(\lambda_0)]]^{(n)}}{n!}=\sum_{j+k=n\atop j,k\geqslant 0}\frac{ \psi_{+\mathrm{R}}^{(k)}(x,t;\lambda_0)\mathbf{B}_j}{j!k!}.
 \end{equation}
 In addition,
\begin{equation}\label{eq:psidd}
    \frac{ \left[\psi_{-\mathrm{R}}(x,t;\bar\lambda_0)\mathrm{adj}[\bar{\mathbf{a}}(\lambda_0)]\right]^{(n)}}{n!}=-\sum_{j+k=n\atop j,k\geqslant 0} \frac{\psi_{+\mathrm{L}}^{(l)}(x,t;\bar\lambda_0)\mathbf{B}_j^\dag}{j!k!}.
\end{equation}
\end{prop}
\begin{proof}
From Eq.\,\eqref{eq:Jostsym1}, it follows   that for $\lambda\in\mathbb{R}$, we have
 \begin{equation}\label{eq:pps0}
     \psi_{\pm\mathrm{L}}^{\dag}(x,t;\bar\lambda)\psi_{\pm\mathrm{R}}(x,t;\lambda)=\mathbf{0}.
  \end{equation}
   This equation can be  analytically continued onto  the upper complex plane $\mathbb{C}^+$. Combining this with Eq.\,\eqref{eq:sS}, we get
  \begin{equation}   \label{mumuid}
     (\psi_{+\mathrm{L}}(\bar\lambda),\psi_{-\mathrm{R}}(\bar\lambda)\bar{\mathbf{a}}^{-1}(\lambda))^\dag
     (\psi_{-\mathrm{L}}(\lambda){\mathbf{a}}^{-1}(\lambda),\psi_{+\mathrm{R}}(\lambda))=\mathbf{I}_4,
  \end{equation}
  for $\lambda\in\mathbb{C}^+\cup\mathbb{R}$.
Moreover,  from  Eqs.\,\eqref{psidet}  and \eqref{eq:Jostjump}, we have for $\lambda\in\mathbb{R}$,
  \begin{equation}
    \det(\psi_{+\mathrm{L}}(\bar\lambda),\psi_{-\mathrm{R}}(\bar\lambda)\bar{\mathbf{a}}^{-1}(\lambda))= \det(\psi_{-\mathrm{L}}(\lambda){\mathbf{a}}^{-1}(\lambda),\psi_{+\mathrm{R}}(\lambda))=1,
  \end{equation}
  which also can be  analytically continued onto  the upper complex plane $\mathbb{C}^+$.

Let us define
\begin{equation}\label{eq:sym3}
  \psi_{-\mathrm{L}}(x,t;\lambda)\mathrm{adj}[\mathbf{a}(\lambda)]=\left(\chi_1(x,t;\lambda),\chi_2(x,t;\lambda)\right).
\end{equation}
By combining Eq.\,\eqref{adjG} with Eq.\,\eqref{mumuid}, we obtain
\begin{equation}\label{eq:detachi}
  \det[\mathbf{a}(\lambda)]\bar{\psi}_{+\mathrm{L}}(\bar\lambda)=\left(-\mathcal{G}[\chi_2(\lambda),\psi_{+3}(\lambda),\psi_{+4}(\lambda)],\mathcal{G}[\chi_1(\lambda),\psi_{+3}(\lambda),\psi_{+4}(\lambda)]\right).
\end{equation}
  We will prove this by induction on $n$.
For the base case $n=0$, it follows from Eq.\,\eqref{eq:detachi} that the vectors $\chi_1(x,t;\lambda_0)$, $\psi_{+3}(x,t;\lambda_0)$  and $\psi_{+4}(x,t;\lambda_0)$ are linearly dependent.  The fact $\mathrm{rank}(\psi_{+\mathrm{R}}(x,t;\lambda))=2$ implies that there must exists a nonzero complex-valued constant vector $\alpha_0$ such that $\chi_1(x,t;\lambda_0)=\psi_{+\mathrm{R}}(x,t;\lambda_0)\alpha_0$.
  Now, assume that the statement is true for some   $0\leqslant n\leqslant j-1$, that is, there exist complex-valued constant vectors $\alpha_0,\alpha_1,\ldots, \alpha_{j-1}$ such  that
    for each $n\in\{0,\ldots,j-1\}$,
 \begin{equation}\label{eq:ls1}
  \frac{ \chi_1^{(n)}(\lambda_0)}{n!}=\sum_{r+s=n\atop r,s\geqslant 0}\frac{ \psi_{+\mathrm{R}}^{(s)}(\lambda_0)\alpha_r}{r!s!},
 \end{equation}
 where the $(x,t)$-dependence is omitted for brevity. We need to show that the statement is also true for $n=j$.
Recalling  $\frac{\partial^k}{\partial \lambda^k}\det[\mathbf{a}(\lambda)]|_{\lambda=\lambda_0}=0$, $k=0,\ldots,j$ and Eq.\,\eqref{eq:detachi}, we find
\begin{equation}\label{eq:ls2}
  \sum_{k+l+s=j\atop k,l,s\geqslant 0}\frac{j!}{k!l!s!}\mathcal{G}\left[\chi_1^{(s)}(\lambda_0),\psi_{+3}^{(k)}(\lambda_0),\psi_{+4}^{(l)}(\lambda_0)\right]=\mathbf{0}.
 \end{equation}

 Substituting Eq.\,\eqref{eq:ls1} into Eq.\,\eqref{eq:ls2} yields
 \begin{align}
   \mathbf{0}= &\mathcal{G}\left[\chi_1^{(j)}(\lambda_0),\psi_{+3}(\lambda_0),\psi_{+4}(\lambda_0)\right]\nonumber\\
   &+\sum_{\substack{k+l+r+s=j\\
  r+s\neq j\\
k,l,r,s\geqslant 0}}\frac{j!}{k!l!r!s!}\mathcal{G}\left[\psi_{+\mathrm{R}}^{(s)}(\lambda_0)\alpha_r,\psi_{+3}^{(k)}(\lambda_0),\psi_{+4}^{(l)}(\lambda_0)\right]\nonumber\\
   =&\mathcal{G}\left[\chi_1^{(j)}(\lambda_0),\psi_{+\mathrm{R}}(\lambda_0)\right]+\Big(\sum_{\substack{ k+l+r+s=j\\
  r+s\neq j\\
  k+r\neq j\\
   l+r\neq j\\
 k,l,r,s\geqslant 0}}+\sum_{\substack{ k+l+r+s=j\\
  r+s\neq j\\
  k+r= j\\
 k,l,r,s\geqslant 0}}+\sum_{\substack{ k+l+r+s=j\\
  r+s\neq j\\
  l+r= j\\
 k,l,r,s\geqslant 0}}\Big)\nonumber\\
 &
  \frac{j!}{k!l!r!s!}\mathcal{G}\left[\psi_{+\mathrm{R}}^{(s)}(\lambda_0)\alpha_r,\psi_{+3}^{(k)}(\lambda_0),\psi_{+4}^{(l)}(\lambda_0)\right]\\
 =&\mathcal{G}\left[\chi_1^{(j)}(\lambda_0),\psi_{+\mathrm{R}}(\lambda_0)\right]+\sum_{\substack{
 k+r=j\\
 k>0,r\geqslant 0
 }}
  \frac{j!}{k!r!}\mathcal{G}\left[\psi_{+\mathrm{R}}(\lambda_0)\alpha_r,\psi_{+3}^{(k)}(\lambda_0),\psi_{+4}(\lambda_0)\right]\nonumber\\
  &+\sum_{\substack{
 l+r=j\\
 l>0,r\geqslant 0
 }}
  \frac{j!}{l!r!}\mathcal{G}\left[\psi_{+\mathrm{R}}(\lambda_0)\alpha_r,\psi_{+3}(\lambda_0),\psi_{+4}^{(l)}(\lambda_0)\right]\nonumber\\
=&\mathcal{G}\left[\chi_1^{(j)}(\lambda_0),\psi_{+\mathrm{R}}(\lambda_0)\right]-\sum_{\substack{
 k+r=j\\
 k>0,r\geqslant 0
 }}
  \frac{j!}{k!r!}\mathcal{G}\left[\psi_{+\mathrm{R}}^{(k)}(\lambda_0)\alpha_r,\psi_{+\mathrm{R}}(\lambda_0)\right]\nonumber\\
  =&\mathcal{G}\bigg[\chi_1^{(j)}(\lambda_0)-\sum_{\substack{
 k+r=j\\
 k>0,r\geqslant 0
 }}
  \frac{j!}{k!r!}\psi_{+\mathrm{R}}^{(k)}(\lambda_0)\alpha_r,\psi_{+\mathrm{R}}(\lambda_0)\bigg].\nonumber
 \end{align}
 Since $\mathrm{rank}(\psi_{+\mathrm{R}}(\lambda_0))=2$,  it follows from Lemma \ref{le:g} that there exists  a constant vector $\alpha_j$ such that
 \begin{equation}
   \chi_1^{(j)}(\lambda_0)-\sum_{\substack{
 k+r=j\\
 k>0,r\geqslant 0
 }}
  \frac{j!}{k!r!}\psi_{+\mathrm{R}}^{(k)}(\lambda_0)\alpha_r=\psi_{+\mathrm{R}}(\lambda_0)\alpha_j,
 \end{equation}
 i.e.,
 \begin{equation}
   \chi_1^{(j)}(\lambda_0)=\sum_{\substack{
 k+r=j\\
 k,r\geqslant 0
 }}
  \frac{j!}{k!r!}\psi_{+\mathrm{R}}^{(k)}(\lambda_0)\alpha_r.
 \end{equation}
 By induction hypothesis, we prove that there exist complex-valued constant vectors $\alpha_0,\alpha_1,\ldots, \alpha_{m}$ such  that
    for each $n\in\{0,\ldots,m\}$,
 \begin{equation}
  \frac{ \chi_1^{(n)}(\lambda_0)}{n!}=\sum_{r+s=n\atop r,s\geqslant 0}\frac{ \psi_{+\mathrm{R}}^{(s)}(\lambda_0)\alpha_r}{r!s!}.
 \end{equation}
 Similarly,  there exist complex-valued constant vectors $\beta_0,\beta_1,\ldots, \beta_{m}$ such  that
    for each $n\in\{0,\ldots,m\}$,
 \begin{equation}
  \frac{ \chi_2^{(n)}(\lambda_0)}{n!}=\sum_{r+s=n\atop r,s\geqslant 0}\frac{ \psi_{+\mathrm{R}}^{(s)}(\lambda_0)\beta_r}{r!s!}.
  \end{equation}
  Let $\mathbf{B}_n=(\alpha_n,\beta_n)$, $n=0,\ldots,m$, we obtain Eq.\,\eqref{eq:psid}.

  As a consequence of Eq.\,\eqref{mumuid},
  \begin{equation}\label{imumuid}
    (\psi_{-\mathrm{L}}(\lambda){\mathbf{a}}^{-1}(\lambda),\psi_{+\mathrm{R}}(\lambda))
    (\psi_{+\mathrm{L}}(\bar\lambda),\psi_{-\mathrm{R}}(\bar\lambda)\bar{\mathbf{a}}^{-1}(\lambda))^\dag=\mathbf{I}_4,\quad \lambda\in\mathbb{C}^+\cup\mathbb{R},
  \end{equation}
  i.e.,
  \begin{equation}\label{iimumuid}
    (\psi_{-\mathrm{L}}(\lambda)\mathrm{adj}[\mathbf{a}(\lambda)],\psi_{+\mathrm{R}}(\lambda))\begin{pmatrix}
      \psi^{\dag}_{+\mathrm{L}}(\bar\lambda)\\
      \mathrm{adj}[\mathbf{a}^{\mathrm{T}}(\lambda)]\psi^{\dag}_{-\mathrm{R}}(\bar\lambda)
    \end{pmatrix}
    =\det[\mathbf{a}(\lambda)]\mathbf{I}_4,\quad \lambda\in\mathbb{C}^+\cup\mathbb{R}.
  \end{equation}
For $n=0,\ldots,m$, differentiating both sides of Eq.\,\eqref{iimumuid} $n$ times with respect to $\lambda$  and evaluating it at $\lambda_0$, we get
\begin{equation}
  \sum_{k+l=n\atop k,l\geqslant 0}\frac{1}{k!l!}(\psi_{-\mathrm{L}}(\lambda_0)\mathrm{adj}[\mathbf{a}(\lambda_0)],\psi_{+\mathrm{R}}(\lambda_0))^{(k)}\begin{pmatrix}
      \psi^{\dag}_{+\mathrm{L}}(\bar\lambda_0)\\
      \mathrm{adj}[\mathbf{a}^{\mathrm{T}}(\lambda_0)]\psi^{\dag}_{-\mathrm{R}}(\bar\lambda_0)
    \end{pmatrix}^{(l)}=\mathbf{0}.
\end{equation}
As $n=0$, combining with Eq.\,\eqref{eq:psid} yields
\begin{equation}
  \psi_{+\mathrm{R}}(\lambda_0)\left[\mathbf{B}_0\psi^{\dag}_{+\mathrm{L}}(\bar\lambda_0)+\mathrm{adj}[\mathbf{a}^{\mathrm{T}}(\lambda_0)]\psi^{\dag}_{-\mathrm{R}}(\bar\lambda_0)\right]=\mathbf{0}.
\end{equation}
Recalling the fact $\mathrm{rank}(\psi_{+\mathrm{R}}(\lambda_0))=2$, we  can deduce that
\begin{equation}
  \mathbf{B}_0\psi^{\dag}_{+\mathrm{L}}(\bar\lambda_0)+\mathrm{adj}[\mathbf{a}^{\mathrm{T}}(\lambda_0)]\psi^{\dag}_{-\mathrm{R}}(\bar\lambda_0)=\mathbf{0},
\end{equation}
which is exactly Eq.\,\eqref{eq:psidd} for $n=0$. By induction, we prove that  Eq.\,\eqref{eq:psidd} holds for $n=0,\ldots,m$.

 Recalling  Eqs.\,\eqref{eq:psid}  and \eqref{eq:psidd}, combining with the symmetry \eqref{eq:Jostsym1}, we find $\mathbf{B}_n^{\mathrm{T}}=\mathbf{B}_n$ for $n=0,\ldots,m$.
\end{proof}

\begin{corollary}\label{cormu}
Assuming that $\lambda_0$ is a zero of $\det[\mathbf{a}(\lambda)]$ with multiplicity $m+1$, we have the following expressions  for each $n\in\{0,\ldots,m\}$:
\begin{align}
  &\frac{ [\mu_{-\mathrm{L}}(x,t;\lambda_0)\mathrm{adj}[\mathbf{a}(\lambda_0)]]^{(n)}}{n!}=\sum_{j+k+l=n\atop j,k,l\geqslant 0} \frac{\Theta^{(k)}(x,t;\lambda_0)\mu_{+\mathrm{R}}^{(l)}(x,t;\lambda_0)\mathbf{B}_j}{j!k!l!},\\
    &\frac{ \left[\mu_{-\mathrm{R}}(x,t;\bar\lambda_0)\mathrm{adj}[\bar{\mathbf{a}}(\lambda_0)]\right]^{(n)}}{n!}=-\sum_{j+k+l=n\atop j,k,l\geqslant 0} \frac{\overline{\Theta^{(k)}(x,t;\lambda_0)}\mu_{+\mathrm{L}}^{(l)}(x,t;\bar\lambda_0)\mathbf{B}_j^\dag}{j!k!l!},\label{eq:mumlpr}
\end{align}
where $\Theta(x,t;\lambda)=\mathrm{e}^{2\mathrm{i}\theta(x,t;\lambda)}$, and  $\mathbf{B}_0,\mathbf{B}_1,\ldots, \mathbf{B}_m$ are given in Proposition \ref{prop:psi}.
\end{corollary}
\begin{proof}
We suppress the $(x,t)$-dependence for simplicity. It follows from Eq.\,\eqref{mJost} and Proposition \ref{prop:psi} that
  \begin{align}
      &\frac{ [\mu_{-\mathrm{L}}(\lambda_0)\mathrm{adj}[\mathbf{a}(\lambda_0)]]^{(n)}}{n!}=\frac{ [\Theta^\frac12 (\lambda_0)\psi_{-\mathrm{L}}(\lambda_0)\mathrm{adj}[\mathbf{a}(\lambda_0)]]^{(n)}}{n!}\nonumber\\
      =&\sum_{r+s=n\atop r,s\geqslant 0}
      \frac{(\Theta^\frac12)^{(r)}(\lambda_0)[\psi_{-\mathrm{L}}(\lambda_0)\mathrm{adj}[\mathbf{a}(\lambda_0)]]^{(s)}}{r!s!}\nonumber\\
      =&\sum_{r+s=n\atop r,s\geqslant 0}\sum_{j+m=s\atop j,m\geqslant 0}\frac{(\Theta^\frac12)^{(r)}(\lambda_0)\psi_{+\mathrm{R}}^{(m)}(\lambda_0)\mathbf{B}_j}{r!j!m!}\nonumber\\
       =&\sum_{r+j+m=n\atop r,j,m\geqslant 0}\frac{(\Theta^\frac12)^{(r)}(\lambda_0)(\Theta^\frac12\mu_{+\mathrm{R}})^{(m)}(\lambda_0)\mathbf{B}_j}{r!j!m!}\\
       =&\sum_{r+j+h+l=n\atop r,j,h,l\geqslant 0}\frac{(\Theta^\frac12)^{(r)}(\lambda_0)(\Theta^\frac12)^{(h)}(\lambda_0)\mu_{+\mathrm{R}}^{(l)}(\lambda_0)\mathbf{B}_j}{r!j!h!l!}\nonumber\\
       =&\sum_{j+k+l=n\atop j,k,l\geqslant 0}\sum_{r+h=k\atop r,h\geqslant 0}\frac{(\Theta^\frac12)^{(r)}(\lambda_0)(\Theta^\frac12)^{(h)}(\lambda_0)}{r!h!}\frac{\mu_{+\mathrm{R}}^{(l)}(\lambda_0)\mathbf{B}_j}{j!l!}\nonumber\\
      =&\sum_{j+k+l=n\atop j,k,l\geqslant 0} \frac{\Theta^{(k)}(\lambda_0)\mu_{+\mathrm{R}}^{(l)}(\lambda_0)\mathbf{B}_j}{j!k!l!}.\nonumber
  \end{align}
Similarly, we can derive Eq.\,\eqref{eq:mumlpr}.
\end{proof}

  Suppose that  $\lambda_0$ is a zero of $\det[\mathbf{a}(\lambda)]$ with multiplicity $m+1$,  then $\frac{1}{\det[\mathbf{a}(\lambda)]}$ has the Laurent series expansion at $\lambda=\lambda_0$,
\begin{equation}
 \frac{1}{\det[\mathbf{a}(\lambda)]}=\frac{a_{-m-1}}{(\lambda-\lambda_0)^{m+1}}+\frac{a_{-m}}{(\lambda-\lambda_0)^{m}}+\cdots+\frac{a_{-1}}{\lambda-\lambda_0}+O(1),\quad \lambda\rightarrow\lambda_0,
\end{equation}
where  $a_{-m-1}\neq 0$ and  $a_{-n-1}=\frac{\tilde{a}^{(m-n)}(\lambda_0)}{(m-n)!}$, $\tilde{a}(\lambda)=\frac{(\lambda-\lambda_0)^{m+1}}{\det[\mathbf{a}(\lambda)]}$, $n=0,\ldots,m$. Combining with Corollary \ref{cormu} yields that, for each $n\in\{0,\ldots,m\}$,
 \begin{align}
      &\underset{\lambda_0}{\res}(\lambda-\lambda_0)^n\mu_{-\mathrm{L}}(x,t;\lambda)\mathbf{a}^{-1}(\lambda)=\sum_{j+k+l+s=m-n\atop j,k,l,s\geqslant 0} \frac{{\tilde{a}}^{(j)}(\lambda_0)\Theta^{(l)}(x,t;\lambda_0)\mu_{+\mathrm{R}}^{(s)}(x,t;\bar\lambda_0)\mathbf{B}_k}{j!k!l!s!},\\
      &\underset{\bar\lambda_0}{\res}(\lambda-\bar\lambda_0)^n\mu_{-\mathrm{R}}(x,t;\lambda)\bar{\mathbf{a}}^{-1}(\bar\lambda)=-\sum_{j+k+l+s=m-n\atop j,k,l,s\geqslant 0} \frac{\overline{\tilde{a}^{(j)}(\lambda_0)\Theta^{(l)}(x,t;\lambda_0)}\mu_{+\mathrm{L}}^{(s)}(x,t;\lambda_0)\bar{\mathbf{B}}_k}{j!k!l!s!}.
 \end{align}
Introduce a symmetric matrix-valued polynomial of degree at most $m$ expressed as:
\begin{equation}\label{eq:f0}
  \mathbf{f}_0(\lambda)=\sum_{h=0}^{m}\sum_{j+k=h\atop j,k\geqslant 0}\frac{\tilde{a}^{(j)}(\lambda_0)\mathbf{B}_k}{j!k!}(\lambda-\lambda_0)^h.
\end{equation}
It is evident that $\mathbf{f}_0(\lambda_0)\neq \mathbf{0}$, therefore,
\begin{equation}\label{eq:resf0}
\begin{split}
  \underset{\lambda_0}{\res}(\lambda-\lambda_0)^n\mu_{-\mathrm{L}}(x,t;\lambda)\mathbf{a}^{-1}(\lambda)=&\sum_{h+l+s=m-n\atop h,l,s\geqslant 0} \frac{\Theta^{(l)}(x,t;\lambda_0)\mu_{+\mathrm{R}}^{(s)}(x,t;\lambda_0)\mathbf{f}_0^{(h)}(\lambda_0)}{h!l!s!}\\
  =&\frac{[\mathrm{e}^{2\mathrm{i}\theta(x,t;\lambda)}\mu_{+\mathrm{R}}(x,t;\lambda)\mathbf{f}_0(\lambda)]^{(m-n)}|_{\lambda=\lambda_0}}{(m-n)!}.
\end{split}
 \end{equation}
 Furthermore,
 \begin{equation}\label{eq:rescf0}
   \underset{\bar\lambda_0}{\res}(\lambda-\bar\lambda_0)^{n}\mu_{-\mathrm{R}}(x,t;\lambda)\bar{\mathbf{a}}^{-1}(\bar\lambda)=-
     \frac{[\mathrm{e}^{-2\mathrm{i}\theta(x,t;\lambda)}\mu_{+\mathrm{L}}(x,t;\lambda)\mathbf{f}_0^\dag(\bar\lambda)]^{(m-n)}|_{\lambda=\bar\lambda_0}}{(m-n)!}.
 \end{equation}
   We shall name $\mathbf{f}_0(\lambda_0),\ldots,\mathbf{f}_0^{(m)}(\lambda_0)$ as the residue constants at the discrete spectrum $\lambda_0$.

 \begin{assume}\label{sing}
   Suppose that the determinant of the matrix  $ \mathbf{a}(\lambda)$ has $N$ zeros, denoted by $\lambda_1,\ldots,\lambda_N $, which are all in the upper half of the complex plane $\mathbb{C}^+$ and are not on the real axis. The zeros have multiplicities  $m_1+1,\ldots,m_N+1$, respectively.
\end{assume}
 \begin{prop}\label{prop:res}
    Under the assumptions of Assumption \ref{sing}, there exists  a unique symmetric matrix-valued polynomial $\mathbf{f}(\lambda)$ of degree less than $\mathcal{N}=\sum_{k=1}^N(m_k+1)$, with the property that $\mathbf{f}(\lambda_k)\neq 0$ such that, for  $k=1,\ldots,N$, $n_k=0,\ldots,m_k$,
   \begin{alignat}{2}
    &\underset{\lambda_k}{\res}(\lambda-\lambda_k)^{n_k}\mu_{-\mathrm{L}}(x,t;\lambda)\mathbf{a}^{-1}(\lambda)
  =\frac{[\mathrm{e}^{2\mathrm{i}\theta(x,t;\lambda)}\mu_{+\mathrm{R}}(x,t;\lambda)\mathbf{f}(\lambda)]^{(m_k-n_k)}|_{\lambda=\lambda_k}}{(m_k-n_k)!},\label{eq:resfa}\\
      & \underset{\bar\lambda_k}{\res}(\lambda-\bar\lambda_k)^{n_k}\mu_{-\mathrm{R}}(x,t;\lambda)\bar{\mathbf{a}}^{-1}(\bar\lambda)=-
     \frac{[\mathrm{e}^{-2\mathrm{i}\theta(x,t;\lambda)}\mu_{+\mathrm{L}}(x,t;\lambda)\mathbf{f}^\dag(\bar\lambda)]^{(m_k-n_k)}|_{\lambda=\bar\lambda_k}}{(m_k-n_k)!}.\label{eq:resfb}
   \end{alignat}
 \end{prop}
 \begin{proof}
 Similar to Eqs.\,\eqref{eq:resf0} and \eqref{eq:rescf0},  there exists for each $k\in\{1,\ldots, N\}$ a symmetric matrix-valued polynomial $\mathbf{f}_k(\lambda)$ of degree no more than $m_k$, where $\mathbf{f}_k(\lambda_k)\neq 0$, such that for $n_k=0,\ldots,m_k$,
 \begin{align}
  &\underset{\lambda_k}{\res}(\lambda-\lambda_k)^{n_k}\mu_{-\mathrm{L}}(x,t;\lambda)\mathbf{a}^{-1}(\lambda)
  =\frac{[\mathrm{e}^{2\mathrm{i}\theta(x,t;\lambda)}\mu_{+\mathrm{R}}(x,t;\lambda)\mathbf{f}_k(\lambda)]^{(m_k-n_k)}|_{\lambda=\lambda_k}}{(m_k-n_k)!},\\
      & \underset{\bar\lambda_k}{\res}(\lambda-\bar\lambda_k)^{n_k}\mu_{-\mathrm{R}}(x,t;\lambda)\bar{\mathbf{a}}^{-1}(\bar\lambda)=-
     \frac{[\mathrm{e}^{-2\mathrm{i}\theta(x,t;\lambda)}\mu_{+\mathrm{L}}(x,t;\lambda)\mathbf{f}_k^\dag(\bar\lambda)]^{(m_k-n_k)}|_{\lambda=\bar\lambda_k}}{(m_k-n_k)!}.
 \end{align}
   Using Hermite interpolation formula, there is a unique  symmetric matrix-valued polynomial $\mathbf{f}(\lambda)$ of degree less than $\displaystyle \mathcal{N}$ such that
    \begin{equation}\label{eq:feqn}
    \begin{cases}
      \mathbf{f}^{(n_1)}(\lambda_1)=\mathbf{f}_1^{(n_1)}(\lambda_1), \quad& n_1=0,\ldots,m_1,\\
     \qquad\qquad \vdots&\\
       \mathbf{f}^{(n_N)}(\lambda_N)=\mathbf{f}_N^{(n_N)}(\lambda_N), \quad& n_N=0,\ldots,m_N.
    \end{cases}
    \end{equation}
    Hence, we have proven  Eqs.\,\eqref{eq:resfa} and \eqref{eq:resfb}.
 \end{proof}
 \begin{remark}
Similar to Eq.\,\eqref{eq:f0}, we observe that the polynomial $\mathbf{f}(\lambda)$ presented in Proposition \ref{prop:res} is uniquely determined by $\det[\mathbf{a}(\lambda)]$,  the collection of $\{\lambda_k,m_k\}_{k=1}^N$ and some constant symmetric matrices $\{\mathbf{B}_{k,0},\ldots,\mathbf{B}_{k,m_k}\}_{k=1}^N$.
Suppose now that a symmetric matrix-valued function $\mathbf{g}(\lambda)$ is analytic at $\lambda_1,\ldots,\lambda_N$. If we replace $\mathbf{f}(\lambda)$ with $\mathbf{f}(\lambda)+\prod_{k=1}^N(\lambda-\lambda_k)^{m_k+1}\mathbf{g}(\lambda)$ in Eq.\,\eqref{eq:feqn}, then it still holds.
  Therefore, we can modify Proposition \ref{prop:res} as follows: ``there exists a symmetric matrix-valued function  $\tilde{\mathbf{f}}(\lambda)$ which is analytic and nonzero at $\lambda_1,\ldots,\lambda_N$ and satisfies Eqs.\,\eqref{eq:resfa} and \eqref{eq:resfb}".
 \end{remark}

\section{Inverse problem}\label{inverse}
In Sect.\,\ref{dp}, we  established the direct scattering map as follows:
\begin{equation}
  \mathcal{D}: u_0(x)\mapsto \left\{\gamma(\lambda),\left\{\lambda_k,m_k, \left\{\mathbf{f}^{(n_k)}(\lambda_k)\right\}_{n_k=0}^{m_k}\right\}_{k=1}^N\right\}.
\end{equation}
In the following, we will discuss the inverse scattering map:
\begin{equation}
  \mathcal{I}: \left\{\gamma(\lambda),\left\{\lambda_k,m_k, \left\{\mathbf{f}^{(n_k)}(\lambda_k)\right\}_{n_k=0}^{m_k}\right\}_{k=1}^N\right\}\mapsto \mathbf{Q}(x,t),
\end{equation}  which can be expressed  in terms of a $4\times 4$  matrix RH problem.

\subsection{ Riemann--Hilbert problem}
Consider a piecewise meromorphic function $\mathbf{M}(x,t;\lambda)$ defined as follows:
  \begin{equation}\label{Mdef}
\begin{aligned}
   \mathbf{M}(x,t;\lambda)&=\left(\mu_{-\mathrm{L}}(x,t;\lambda)\mathbf{a}^{-1}(\lambda),\mu_{+\mathrm{R}}(x,t;\lambda)\right),\quad &&\lambda\in \mathbb{C}^+,\\
  \mathbf{M}(x,t;\lambda)&=\left(\mu_{+\mathrm{L}}(x,t;\lambda),\mu_{-\mathrm{R}}(x,t;\lambda)\bar{\mathbf{a}}^{-1}(\bar\lambda)\right),\quad &&\lambda\in \mathbb{C}^-.
\end{aligned}
 \end{equation}
We can use this to formulate  a  $4\times 4$   matrix RH problem:
\begin{problem}\label{frhp}
Find a $4\times 4$ matrix-valued function $\mathbf{M}(x,t;\lambda)$ that satisfies the given properties:
\begin{itemize}
  \item \textbf{Analyticity:} $\mathbf{M}(x,t;\lambda)$ is a $4\times 4$ matrix-valued function that is analytic in
 $\lambda$ for $\lambda\in\mathbb{C}\backslash\left(\mathbb{R}\cup\{\lambda_k,\bar\lambda_k\}_{k=1}^N\right)$;
  \item \textbf{Residues:} For $ k=1,\ldots,N$, at $\lambda=\lambda_k,\bar\lambda_k$, $\mathbf{M}(x,t;\lambda)$ has  poles of order $m_k+1$ and the residues satisfy the following conditions:
\begin{subequations}\label{res1}
  \begin{align}
    &\underset{\lambda_k}{\res} {(\lambda-\lambda_k)^{n_k}\mathbf M}(x,t;\lambda)=\bigg(\frac{[\mathrm{e}^{2\mathrm{i}\theta(x,t;\lambda)}\mathbf{M}_{\mathrm{R}}(x,t;\lambda)\mathbf{f}(\lambda)]^{(m_k-n_k)}|_{\lambda=\lambda_k}}{(m_k-n_k)!},\mathbf{0}\bigg),  \\
    &\underset{\bar\lambda_k}{\res} {(\lambda-\bar\lambda_k)^{n_k}\mathbf M}(x,t;\lambda)=\bigg(\mathbf{0},-\frac{[\mathrm{e}^{-2\mathrm{i}\theta(x,t;\lambda)}\mathbf{M}_{\mathrm{L}}(x,t;\lambda)\mathbf{f}^\dag(\bar\lambda)]^{(m_k-n_k)}|_{\lambda=\bar\lambda_k}}{(m_k-n_k)!}\bigg),
  \end{align}
\end{subequations}
for  $n_k=0,\ldots, m_k$;
  \item \textbf{Jump:} The values of $\mathbf{M}(x,t;\lambda)$ on either side of the real axis has a ``jump" relation:
  \begin{equation}\label{Mjump}
  \mathbf{M}_+(x,t;\lambda)=\mathbf{M}_-(x,t;\lambda)\mathbf{J}(x,t;\lambda),\quad \lambda\in\mathbb{R},
\end{equation}
where $ \mathbf{M}_{\pm}(x,t;\lambda)=\displaystyle\lim_{\epsilon\downarrow 0}\mathbf{M}(x,t;\lambda\pm\mathrm{i}\epsilon)$,   and the jump matrix can be expressed in terms of the reflection coefficient $\gamma(\lambda)$ as follows:
\begin{equation}\label{jump}
  \mathbf{J}(x,t;\lambda)=\begin{pmatrix}
    \mathbf{I}_2+\bar{\gamma}(\bar\lambda)\gamma(\lambda)&\mathrm{e}^{-2\mathrm{i}\theta(x,t;\lambda)}\bar{\gamma}(\bar\lambda)\\
    \mathrm{e}^{2\mathrm{i}\theta(x,t;\lambda)}\gamma(\lambda)&\mathbf{I}_2
  \end{pmatrix};
\end{equation}
  \item \textbf{Normalization:}  At infinity $\mathbf{M}(x,t;\lambda)$ has the form:
\begin{equation}\label{eq:masy}
\mathbf{M}(x,t;\lambda)=\mathbf{I}_4+O(\lambda^{-1}),\quad \lambda\in\mathbb{C}\backslash \mathbb{R}\rightarrow\infty.
\end{equation}

\end{itemize}
\end{problem}
By following a similar analysis as Ref.\,\cite{FI1996}, we can map the above problem in the RH framework to a regular problem that has no singularities. The unique solution to the regular RH problem \ref{frhp} only exists if a vanishing lemma exists, as described below (refer to Refs.\,\cite{AF203,Zhou1989}).
\begin{lemma}{(Vanishing Lemma)}\label{vanish}
The regular RH problem for $\mathbf{M}(x,t;\lambda)$ obtained from   RH problem \ref{frhp} by replacing the asymptotics \eqref{eq:masy} with
\begin{equation}
 \mathbf{M}(x,t;\lambda)=O(\lambda^{-1}), \qquad \lambda\rightarrow \infty,
\end{equation}
 has only the zero solution.
\end{lemma}

\subsection{Reconstruction of potential }

\begin{theorem}(Reconstruction formula)\label{Thinv}
Suppose $\mathbf{M}(x,t;\lambda)$ is the solution to RH problem \ref{frhp}. Then,
\begin{equation}\label{recon1}
   \mathbf{Q}(x,t)= 2\mathrm{i}\lim_{\lambda\rightarrow \infty}\lambda\mathbf{M}_{\mathrm{UR}}(x,t;\lambda),
\end{equation}
solves the  spin-1 GP equation \eqref{Eq:SGPE}.
\end{theorem}
\begin{proof}
  This statement can be proven using the dressing method, as described in Ref.\,\cite{Fokas2008}.
\end{proof}

 In fact, the RH problem \ref{frhp} can be regularized by subtracting any pole contributions and the leading order asymptotics at infinity:
\begin{equation}
 \begin{split}
        \mathcal{M}(x,t;\lambda)=&\mathbf{M}(x,t;\lambda)-\mathbf{I}_4\\
        &-\sum_{k=1}^{N}\sum_{n_k=0}^{m_k}\left(\frac{\underset{\lambda_k}{\res}(\lambda-\lambda_k)^{n_k}\mathbf{M}(x,t;\lambda)}{(\lambda-\lambda_k)^{n_k+1}}
        +
        \frac{\underset{\bar \lambda_k}{\res}(\lambda-\bar\lambda_k)^{n_k}\mathbf{M}(x,t;\lambda)}{(\lambda-\bar\lambda_k)^{n_k+1}}\right).
 \end{split}
 \end{equation}
Consequently, the piecewise holomorphic function $\mathcal{M}(x,t;\lambda)$ satisfies
\begin{align}
  &  \mathcal{M}_+(x,t;\lambda)-\mathcal{M}_-(x,t;\lambda)=\mathbf{M}_-(x,t;\lambda)(\mathbf{J}(x,t;\lambda)-\mathbf{I}_4),\quad &&\lambda\in\mathbb{R},\\
  &\mathcal{M}(x,t;\lambda)\rightarrow \mathbf{0},\quad&& \lambda\rightarrow\infty.
\end{align}
Using Sokhotski--Plemelj formula, we can express $\mathcal{M}(x,t;\lambda)$ as an integral over the real line:
 \begin{equation}
      \mathcal{M}(x,t;\lambda)=\frac{1}{2\pi\mathrm{i}}\int_{\mathbb{R}} \frac{\mathbf{M}_-(x,t;\zeta)(\mathbf{J}(x,t;\zeta)-\mathbf{I}_4)}{\zeta-\lambda}\mathrm{d}\zeta.
 \end{equation}
Alternatively, we can solve the RH problem \ref{frhp} using the system of algebraic-integral equations
 \begin{subequations}
 \begin{align}
    \mathbf{M}_{\mathrm{L}}(x,t;\lambda)=&\begin{pmatrix}
      \mathbf{I}_2\\\mathbf{0}
    \end{pmatrix}+\sum_{k=1}^N\sum_{n_k=0}^{m_k}\frac{[\mathrm{e}^{2\mathrm{i}\theta(x,t;\lambda)}\mathbf{M}_{\mathrm{R}}(x,t;\lambda)\mathbf{f}(\lambda)]^{(m_k-n_k)}|_{\lambda=\lambda_k}}{(\lambda-\lambda_k)^{n_k+1}(m_k-n_k)!}\nonumber\\
     &+\frac{1}{2\pi\mathrm{i}}\int_{\mathbb{R}} \frac{\mathbf{M}_-(x,t;\zeta)(\mathbf{J}(x,t;\zeta)-\mathbf{I}_4)_{\mathrm{L}}}{\zeta-\lambda}\mathrm{d}\zeta,\\
 \mathbf{M}_{\mathrm{R}}(x,t;\lambda)=&\begin{pmatrix}
  \mathbf{0}\\\mathbf{I}_2
 \end{pmatrix}-\sum_{k=1}^N\sum_{n_k=0}^{m_k}\frac{[\mathrm{e}^{-2\mathrm{i}\theta(x,t;\lambda)}\mathbf{M}_{\mathrm{L}}(x,t;\lambda)\mathbf{f}^\dag(\bar\lambda)]^{(m_k-n_k)}|_{\lambda=\bar\lambda_k}}{(\lambda-\bar\lambda_k)^{n_k+1}(m_k-n_k)!}\nonumber\\
 &+\frac{1}{2\pi\mathrm{i}}\int_{\mathbb{R}} \frac{\mathbf{M}_-(x,t;\zeta)(\mathbf{J}(x,t;\zeta)-\mathbf{I}_4)_{\mathrm{R}}}{\zeta-\lambda}\mathrm{d}\zeta.
 \end{align}
 \end{subequations}

 \begin{corollary}
   The solution to the   spin-1 GP equation \eqref{Eq:SGPE}  has been reconstructed as follows:
   \begin{equation}\label{recon}
   \begin{split}
      \mathbf{Q}(x,t)=&-2\mathrm{i}\sum_{k=1}^{N}\frac{[\mathrm{e}^{-2\mathrm{i}\theta(x,t;\lambda)}\mathbf{M}_{\mathrm{UL}}(x,t;\lambda)\mathbf{f}^\dag(\bar\lambda)]^{(m_k)}|_{\lambda=\bar\lambda_k}}{m_k!}\\
 &+\frac{1}{\pi }\int_{\mathbb{R}} \left[\mathbf{M}_-(x,t;\zeta)(\mathbf{J}(x,t;\zeta)-\mathbf{I}_4)\right]_{\mathrm{UR}}\mathrm{d}\zeta.
   \end{split}
   \end{equation}
 \end{corollary}

 \subsection{Reflectionless potential}\label{less}
 We can now explicitly reconstruct the potential $ \mathbf{Q}(x,t)$ in the reflectionless case which means that $\gamma(\lambda)=\mathbf{0}$ for $\lambda\in\mathbb{R}$. Because of this, there is no jump across the contour  $\mathbb{R}$  and the inverse problem can be reduced to an algebraic system
  \begin{subequations}
 \begin{align}
    \mathbf{M}_{\mathrm{UL}}(\lambda)
    =&\mathbf{I}_2+\sum_{k=1}^N\frac{1}{m_k!}\partial^{m_k}_\mu\left.\left[\frac{\mathrm{e}^{2\mathrm{i}\theta(\mu)}\mathbf{M}_{\mathrm{UR}}(\mu)\mathbf{f}(\mu)}{\lambda-\mu}\right]\right|_{\mu=\lambda_k},\label{algeM1}\\
 \mathbf{M}_{\mathrm{UR}}(\mu)=&-\sum_{l=1}^N\frac{1}{m_l!}\partial_\nu^{m_l}\left.\left[\frac{\mathrm{e}^{-2\mathrm{i}\theta(\nu)}\mathbf{M}_{\mathrm{UL}}(\nu)\mathbf{f}^\dag(\bar\nu)}{\mu-\nu}\right]\right|_{\nu=\bar\lambda_l},\label{algeM2}
 \end{align}
 \end{subequations}
 where the $(x,t)$-dependence is omitted for brevity.

Let
\begin{align}
  \mathbf{h}(\lambda)=&\mathrm{e}^{2\mathrm{i}\theta(\lambda)}\mathbf{f}(\lambda),\quad \mathbf{F}(\lambda)= -2\mathrm{i}\mathbf{M}_{\mathrm{UL}}(\lambda)\mathbf{h}^\dag(\bar\lambda),\\
  \mathbf{G}(\lambda)=& \mathbf{F}(\lambda)+2\mathrm{i}\mathbf{h}^\dag(\bar\lambda)+\sum_{k=1}^N\sum_{l=1}^N\frac{\partial_{\mu}^{m_k}\partial_{\nu}^{m_l}}{m_k!m_l!}\left.\left(\frac{\mathbf{F}(\nu)
    \mathbf{h}(\mu)\mathbf{h}^\dag(\bar\lambda)}{(\lambda-\mu)(\mu-\nu)}\right)\right|_{\mu=\lambda_k\atop \nu=\bar\lambda_l}.\label{eq:G}
\end{align}
  Substituting Eq.\,\eqref{algeM2} into  Eq.\,\eqref{algeM1}, we find
  \begin{equation}
  \mathbf{G}(x,t;\lambda)=\mathbf{0},\quad (x,t;\lambda)\in\mathbb{R}^2\times\mathbb{C}\ \mbox{and}\ \lambda\neq  \lambda_1,\ldots, \lambda_N.
\end{equation}

\begin{theorem}\label{refthe1}
   In the reflectionless case, the solution to the spin-1 GP equation \eqref{Eq:SGPE} can be expressed  as follows:
   \begin{equation}
      \mathbf{Q}(x,t)= \sum_{k=1}^N\frac{\mathbf{F}^{(m_k)}(x,t;\bar\lambda_k)}{m_k!},
   \end{equation}
   where $\{\mathbf{F}^{(j_k)}(x,t;\bar\lambda_k)\}_{k=1,\ldots, N\atop  j_k=0,\ldots,m_k}$ is the solution to the following algebraic system
    \begin{equation}\label{algesys}
     \begin{cases}
      \frac{ \mathbf{G}^{(j_1)}(x,t;\bar\lambda_1)}{j_1!}=\mathbf{0},\quad &j_1=0,\ldots,m_1,\\
      \qquad \qquad\vdots\\
        \frac{\mathbf{G}^{(j_N)}(x,t;\bar\lambda_N)}{j_N!}=\mathbf{0},\quad &j_N=0,\ldots,m_N.
     \end{cases}
   \end{equation}
\end{theorem}
It is necessary to verify the existence and uniqueness of the solution to the algebraic system \eqref{algesys}. The following section will address this matter.
\section{Existence and uniqueness}\label{sect:regular}
Let $\mathcal{N}=2\sum_{k=1}^N(m_k+1)$. Then we can write the transpose of Eq.\,\eqref{eq:G}:
\begin{align}
  \mathbf{G}^{\mathrm{T}}(x,t;\lambda)=& \mathbf{F}^{\mathrm{T}}(x,t;\lambda)+2\mathrm{i}\mathbf{h}^\dag(x,t;\bar\lambda)\nonumber\\
  &+\sum_{k=1}^N\sum_{l=1}^N\frac{\partial_{\mu}^{m_k}\partial_{\nu}^{m_l}}{m_k!m_l!}
  \left.\left(\frac{\mathbf{h}^\dag(x,t;\bar\lambda)
    \mathbf{h}(x,t;\mu)\mathbf{F}^{\mathrm{T}}(x,t;\nu)}{(\lambda-\mu)(\mu-\nu)}\right)\right|_{\mu=\lambda_k\atop \nu=\bar\lambda_l}.
\end{align}
Since the potential matrix $\mathbf{Q}(x,t)$ is symmetric, Theorem\,\ref{refthe1} is equivalent to the following form.
\begin{theorem}\label{refthe}
   In the reflectionless case, the solution to the spin-1 GP equation \eqref{Eq:SGPE} can be expressed  as follows:
   \begin{equation}\label{qcons}
      \mathbf{Q}(x,t)=\sum_{k=1}^N\frac{{\mathbf{F}^{\mathrm{T}}}^{(m_k)}(x,t;\bar\lambda_k)}{m_k!},
   \end{equation}
   where $\{{{\mathbf{F}}^{\mathrm{T}}}^{(j_k)}(x,t;\bar\lambda_k)\}_{k=1,\ldots, N\atop  j_k=0,\ldots,m_k}$ is the solution to the following algebraic system
    \begin{equation}
     \begin{cases}
      \frac{ {\mathbf{G}^{\mathrm{T}}}^{(j_1)}(x,t;\bar\lambda_1)}{j_1!}=\mathbf{0},\quad &j_1=0,\ldots,m_1,\\
      \qquad \qquad\vdots\\
        \frac{{\mathbf{G}^{\mathrm{T}}}^{(j_N)}(x,t;\bar\lambda_N)}{j_N!}=\mathbf{0},\quad &j_N=0,\ldots,m_N.
     \end{cases}
   \end{equation}
\end{theorem}

  In the following, we will  rigorously prove Theorem \ref{refthe}. Before doing so, we will present some preliminary statements.
\begin{definition}
  For a $\mathcal{N}\times \mathcal{N}$ matrix $\mathbf{X}$, let
  \[\mathbf{X}=\begin{pmatrix}
                     \mathbf{X}_{11}&\mathbf{X}_{12}&\cdots\mathbf{X}_{1N}\\
                     \mathbf{X}_{21}&\mathbf{X}_{22}&\cdots\mathbf{X}_{2N}\\
                     &\vdots&\\
                     \mathbf{X}_{N1}&\mathbf{X}_{N2}&\cdots\mathbf{X}_{NN}\\

                    \end{pmatrix},\quad \mathbf{X}_{jk}=\begin{pmatrix}
                     \mathbf{X}_{jk}^{11}&\mathbf{X}_{jk}^{12}&\cdots\mathbf{X}_{jk}^{1,m_k+1}\\
                     \mathbf{X}_{jk}^{21}&\mathbf{X}_{jk}^{22}&\cdots\mathbf{X}_{jk}^{2,m_k+1}\\
                     &\vdots&\\
                     \mathbf{X}_{jk}^{m_j+1,1}&\mathbf{X}_{jk}^{m_j+1,2}&\cdots\mathbf{X}_{jk}^{m_j+1,m_k+1}\\

                    \end{pmatrix},\]
                    where $\mathbf{X}_{jk}$ is a $(2m_j+2)\times(2m_k+2)$ matrix for $j,k\in\{1,\ldots,N\}$,
                    $\mathbf{X}_{jk}^{rs}$ is a $2\times 2$ matrix for  $r\in\{1,\ldots,m_j+1\}$, $s\in\{1,\ldots,m_k+1\}$, we call  $\mathbf{X}_{jk}^{rs}$  the $(r_j,s_k)$-entry of $\mathbf{X}$.
\end{definition}
\begin{definition}
 Define a matrix-valued function of $(x,t)$:
  \begin{equation}\label{Hdef}
  \mathbf{H}=\mathbf{H}(x,t)=-2\mathrm{i}\left(\frac{  \mathbf{h}^{(0)}(\lambda_1 )}{0!},\ldots,\frac{\mathbf{h}^{(m_1)}(\lambda_1 )}{m_1!},\ldots,\frac{\mathbf{h}^{(0)}(\lambda_N ) }{0!},\ldots,\frac{\mathbf{h}^{(m_N)}(\lambda_N ) }{m_N!}\right)^{\dag}.
 \end{equation}
 \end{definition}
 \begin{definition}
Define two $\mathcal{N}\times \mathcal{N}$  matrix-valued functions of $(x,t)$:
 \begin{equation}\label{ABdef}
 \mathbf{A}=\mathbf{A}(x,t),\quad \mathbf{B}=\mathbf{B}(x,t),
 \end{equation}
  with $(r_j, s_k)$-entry for $j,k=1,\ldots,N$, $r=1,\ldots,m_j+1$, $s=1,\ldots,m_k+1$, respectively,
 \begin{equation}\label{AB}
   \mathbf{A}_{jk}^{rs}=-\mathrm{i}\frac{\partial_{\lambda}^{(r-1)}\partial_\mu^{(s-1)}}{(r-1)!(s-1)!}\left.\left(\frac{\mathbf{h}^\dag(\bar\lambda)}{\lambda-\mu}\right)\right|_{\lambda=\bar\lambda_j\atop \mu=\lambda_k},\quad \mathbf{B}_{jk}^{rs}=\mathbf{A}_{jk}^{m_j+2-r,m_k+2-s}.
 \end{equation}
\end{definition}

Therefore,   there exists an equivalent statement to Theorem \ref{refthe} as follows.
\begin{theorem}\label{Th:alge2}
 For any fixed $(x,t)\in\mathbb{R}^2$, the algebraic system
 \begin{equation}\label{alge}
   \left(\mathbf{I}_\mathcal{N}+\mathbf{A}(x,t)\bar{\mathbf{B}}(x,t)\right)\tilde{\mathbf{F}}(x,t)=\mathbf{H}(x,t),
 \end{equation}
with the unknown column vector
 \begin{equation}\label{Fdef}
   \tilde{\mathbf{F}}=\tilde{\mathbf{F}}(x,t)=\left(\frac{\mathbf{F}^{(0)}(\bar\lambda_1)}{0!},\ldots,\frac{\mathbf{F}^{(m_1)}(\bar\lambda_1)}{m_1!},\ldots,\frac{\mathbf{F}^{(0)}(\bar\lambda_N)}{0!},\ldots,\frac{\mathbf{F}^{(m_N)}(\bar\lambda_N)}{m_N!}\right)^{\mathrm{T}},
 \end{equation}
 has a unique solution, and the solution to the  spin-1 GP equation \eqref{Eq:SGPE} can be reconstructed as
  \begin{equation}\label{eq:qre2}
   \mathbf{Q}(x,t)=\alpha\left(\mathbf{I}_\mathcal{N}+\mathbf{A}(x,t)\bar{\mathbf{B}}(x,t)\right)^{-1}\mathbf{H}(x,t),
 \end{equation}
 with \begin{equation}\label{alphadef}
   \alpha=(\mathbf{0}_{2\times 2m_1},\mathbf{I}_2,\mathbf{0}_{2\times 2m_2},\mathbf{I}_2,\ldots,\mathbf{0}_{2\times 2m_N},\mathbf{I}_2).
 \end{equation}
\end{theorem}

 \begin{definition}
   Define a symmetric   real-valued matrix
 \begin{equation}\label{Gadef}
\mathbf{\Gamma}=\begin{pmatrix}
         \mathbf{\Gamma}_1\otimes \mathbf{I}_2 &&& \\
         &\mathbf{\Gamma}_2\otimes \mathbf{I}_2&& \\
         &&\ddots& \\
         &&&\mathbf{\Gamma}_N\otimes \mathbf{I}_2
       \end{pmatrix}, \end{equation}
       where $\mathbf{\Gamma}_j=\begin{pmatrix}
  &&&1\\
  &&1&\\
  &\vdots&&\\
  1&&&
\end{pmatrix}_{(m_j+1)\times (m_j+1)}$ for $ j=1,\ldots, N$.

 \end{definition}
It is evident that $\mathbf{\Gamma}^2=\mathbf{I}_\mathcal{N}$, and we can derive from Eq.\,\eqref{AB} that
 \begin{equation}\label{ABG}
   \mathbf{B}(x,t)=\mathbf{\Gamma}\mathbf{A}(x,t)\mathbf{\Gamma}.
 \end{equation}
 \begin{definition}
 Define a  $\mathcal{N}\times \mathcal{N}$ matrix-valued function  $
   \hat{\mathbf{A}}(x,t)$   with $(r_j, s_k)$-entry,
 \begin{equation}\label{Ahdef}
   \hat{\mathbf{A}}_{jk}^{rs}=\begin{cases}
     \frac{[\mathbf{h}^{(r-s)}(x,t;\lambda_j)]^\dag}{(r-s)!},\quad &j=k\ \mbox{and} \ s\leqslant r,\\
     \mathbf{0},&\mbox{otherwise},
   \end{cases}
 \end{equation}
 and  a  $\mathcal{N}\times \mathcal{N}$ matrix $ \tilde{\mathbf{A}}$  with $(r_j, s_k)$-entry,
 \begin{equation}\label{Atdef}
   \tilde{\mathbf{A}}_{jk}^{rs}=-\mathrm{i}\frac{\partial_{\lambda}^{(r-1)}\partial_\mu^{(s-1)}}{(r-1)!(s-1)!}\left.\left(\frac{1}{\lambda-\mu}\right)\right|_{\lambda=\bar\lambda_j\atop \mu=\lambda_k}\mathbf{I}_2,
 \end{equation}
 for $j,k=1,\ldots,N$, $r=1,\ldots,m_j+1$, $s=1,\ldots,m_k+1$.
 \end{definition}
 Indeed,  upon performing direct calculations, we have discovered
 \begin{equation}\label{AA}
   \hat{\mathbf{A}}(x,t)\tilde{\mathbf{A}}=\mathbf{A}(x,t).
 \end{equation}

 \begin{prop}\label{pr:ta}
  The  Hermitian matrix $\tilde{\mathbf{A}}$ defined by Eq.\,\eqref{Atdef} is    positive definite.
 \end{prop}
  \begin{proof}
For $j,k=1,\ldots,N$, $r=1,\ldots,m_j+1$, $s=1,\ldots,m_k+1$.
    Denote
   \begin{equation}
     h_{r_j}(y)=\overline{\left.\frac{\partial^{(r-1)}_\lambda(\mathrm{e}^{\mathrm{i}\lambda y})}{(r-1)!}\right|_{\lambda=\lambda_j}}=\left.\frac{\partial^{(r-1)}_\lambda(\mathrm{e}^{-\mathrm{i}\lambda y})}{(r-1)!}\right|_{\lambda=\bar\lambda_j}=\frac{(-\mathrm{i}y)^{r-1}\mathrm{e}^{-\mathrm{i}\bar\lambda_j y}}{(r-1)!}.
   \end{equation}
   Noting that
$   |h_{r_j}(y)|^2=\frac{\mathrm{e}^{-2\mathrm{Im}(\lambda_j)y}|y|^{2r-2}}{[(r-1)!]^2}$,
  and recalling  that $\lambda_j\in\mathbb{C}^+$, we conclude  that $h_{r_j}(y)\in L^2(\mathbb{R}^+)$.
  Introduce an inner product $\langle\cdot,\cdot\rangle$:
     \begin{equation}
     \begin{split}
        \langle h_{r_j}(y),h_{s_k}(y)\rangle=&\int_{0}^{+\infty}h_{r_j}(y)\overline{h_{s_k}(y)}\mathrm{d}y\\
          =&\int_{0}^{+\infty} \left.\frac{\partial^{(r-1)}_\lambda(\mathrm{e}^{-\mathrm{i}\lambda y})}{(r-1)!}\right|_{\lambda=\bar\lambda_j} \left.\frac{\partial^{(s-1)}_\mu(\mathrm{e}^{\mathrm{i}\mu y})}{(s-1)!}\right|_{\mu=\lambda_k}\mathrm{d}y\\
          =&\left.\frac{\partial^{(r-1)}_\lambda\partial^{(s-1)}_\mu}{(r-1)!(s-1)!}\left(
          \int_{0}^{+\infty}\mathrm{e}^{\mathrm{i}(\mu-\lambda) y}\mathrm{d}y\right)\right|_{\lambda=\bar\lambda_j\atop \mu=\lambda_k}\\
          =&-\mathrm{i}\frac{\partial_{\lambda}^{(r-1)}\partial_\mu^{(s-1)}}{(r-1)!(s-1)!}\left.\left(\frac{1}{\lambda-\mu}\right)\right|_{\lambda=\bar\lambda_j\atop \mu=\lambda_k}.
     \end{split}
   \end{equation}
   Let
   \begin{equation}
     \mathbf{C}=\begin{pmatrix}
                  \mathbf{C}_{11}&\mathbf{C}_{12}&\cdots\mathbf{C}_{1N}\\
                     \mathbf{C}_{21}&\mathbf{C}_{22}&\cdots\mathbf{C}_{2N}\\
                     &\vdots&\\
                     \mathbf{C}_{N1}&\mathbf{C}_{N2}&\cdots\mathbf{C}_{NN}
                \end{pmatrix}, \quad \mathbf{C}_{jk}=( \langle h_{r_j},h_{s_k}\rangle)_{(m_j+1)\times(m_k+1)}.
   \end{equation}Therefore, $\mathbf{C}$  is a well-known Gram matrix that is Hermitian, positive semi-definite. More precisely, $\mathbf{C}$ is positive definite, this because $\lambda_1,\ldots,\lambda_N$ are distinct and the
functions
  \begin{equation}
     \mathrm{e}^{-\mathrm{i}\bar\lambda_1 y},y\mathrm{e}^{-\mathrm{i}\bar\lambda_1 y},\ldots,y^{m_1}\mathrm{e}^{-\mathrm{i}\bar\lambda_1 y},\ldots,
      \mathrm{e}^{-\mathrm{i}\bar\lambda_N y},y\mathrm{e}^{-\mathrm{i}\bar\lambda_N y},\ldots,y^{m_N}\mathrm{e}^{-\mathrm{i}\bar\lambda_N y},
   \end{equation}
   are linearly independent for $y\in\mathbb{R}^+$.

   Based on the equation $\tilde{\mathbf{A}}_{jk}^{rs}=\langle h_{r_j},h_{s_k}\rangle\mathbf{I}_2$, it follows that $\tilde{\mathbf{A}}=\mathbf{C}\otimes \mathbf{I}_2$. As a result, $\tilde{\mathbf{A}}$ is  a positive definite Hermitian matrix.
   \end{proof}

 \begin{prop}\label{GaL}
  For any fixed $(x,t)\in\mathbb{R}^2$,
 \begin{equation}\label{GaA}
  \mathbf{\Gamma}\hat{\mathbf{A}}(x,t)=\hat{\mathbf{A}}^{\mathrm{T}}(x,t)\mathbf{\Gamma},\quad \mathbf{\Gamma}\bar{\hat{\mathbf{A}}}(x,t)=\hat{\mathbf{A}}^\dag(x,t)\mathbf{\Gamma}.
 \end{equation}
 \end{prop}
 \begin{proof}
 Direct verification is possible.
 \end{proof}

 \begin{prop}
     The solution to the algebraic system \eqref{alge} exists uniquely.
   \end{prop}
\begin{proof}
 From Eqs.\,\eqref{AA} and \eqref{GaA}, we can deduce  that
\begin{equation}
  \mathbf{A}(x,t)\bar{\mathbf{B}}(x,t)=\mathbf{A}(x,t)\mathbf{\Gamma}\bar{\mathbf{A}}(x,t)\mathbf{\Gamma}
  =\hat{\mathbf{A}}(x,t)\tilde{\mathbf{A}}\mathbf{\Gamma}\bar{\hat{\mathbf{A}}}(x,t)\bar{\tilde{\mathbf{A}}}\mathbf{\Gamma}
  =\hat{\mathbf{A}}(x,t)\tilde{\mathbf{A}}\hat{\mathbf{A}}^\dag(x,t)\mathbf{\Gamma} \bar{\tilde{\mathbf{A}}}\mathbf{\Gamma}.
\end{equation}
Combining Eq.\,\eqref{Gadef} with Proposition \ref{pr:ta}, we can conclude that both     $\hat{\mathbf{A}}(x,t)\tilde{\mathbf{A}}\hat{\mathbf{A}}^\dag(x,t)$ and $\mathbf{\Gamma} \bar{\tilde{\mathbf{A}}}\mathbf{\Gamma}$ are  Hermitian  positive definite matrices for any fixed $(x,t)\in\mathbb{R}^2$. Therefore, there must exist  a Hermitian positive definite matrix $\mathbf{C}$ such that $\mathbf{C}^2=\mathbf{\Gamma} \bar{\tilde{\mathbf{A}}}\mathbf{\Gamma}$. It is clear that $\mathbf{C}\hat{\mathbf{A}}(x,t)\tilde{\mathbf{A}}\hat{\mathbf{A}}^\dag(x,t)\mathbf{C}$ is also a Hermitian positive definite matrix for any fixed $(x,t)\in\mathbb{R}^2$. Therefore,
   \begin{equation}
   \begin{split}
      \det \left(\mathbf{I}_\mathcal{N}+\mathbf{A}(x,t)\bar{\mathbf{B}}(x,t)\right)=&  \det \left(\mathbf{I}_\mathcal{N}+\hat{\mathbf{A}}(x,t)\tilde{\mathbf{A}}\hat{\mathbf{A}}^\dag(x,t)\mathbf{C}^2\right)\\
      =&\det \left(\mathbf{I}_\mathcal{N}+\mathbf{C}\hat{\mathbf{A}}(x,t)\tilde{\mathbf{A}}\hat{\mathbf{A}}^\dag(x,t)\mathbf{C}\right)>1.
   \end{split}
   \end{equation}
   Using Cramer's Rule, we can now prove the existence and uniqueness of the solution to the algebraic system \eqref{alge}.
\end{proof}

\section{Some explicit solutions}\label{sect:explicit}
\begin{figure}[ht]
\centering
\begin{subfigure}{\textwidth}
\centering
\includegraphics[scale=.5]{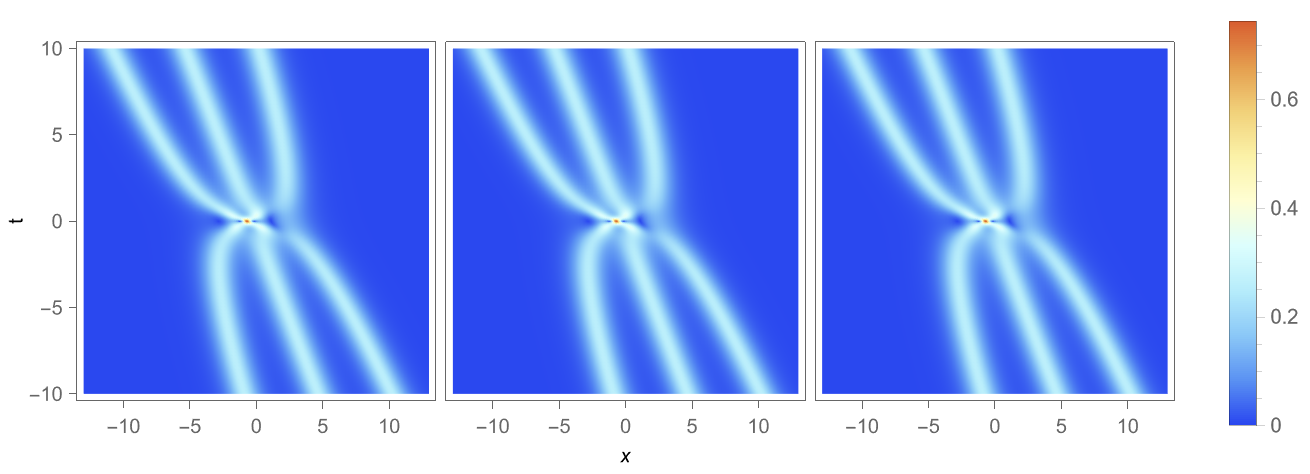}
\end{subfigure}
\begin{subfigure}{\textwidth}
\centering
\includegraphics[scale=.5]{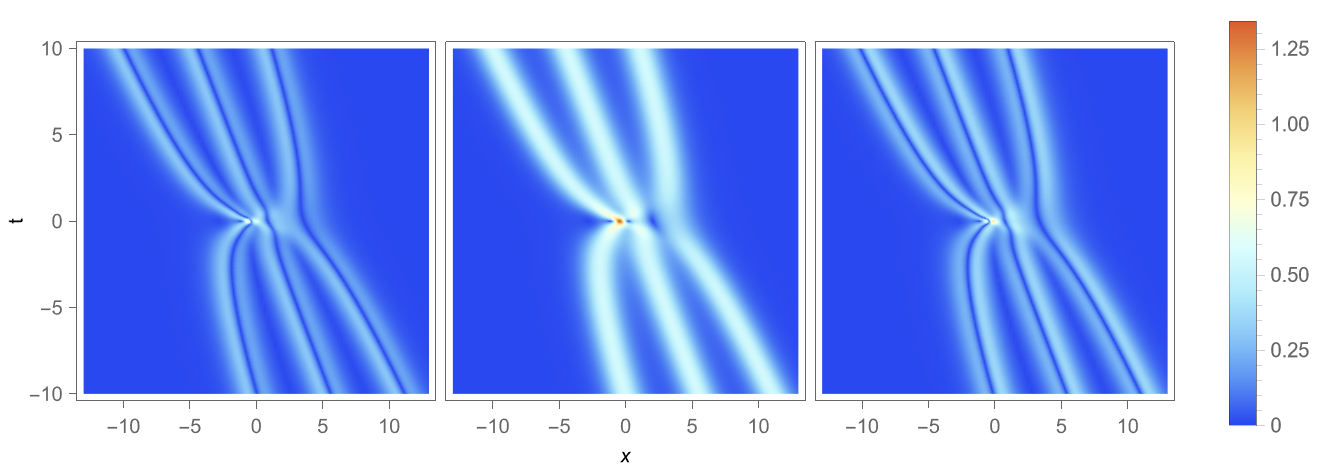}
\end{subfigure}
\begin{subfigure}{\textwidth}
\centering
\includegraphics[scale=.5]{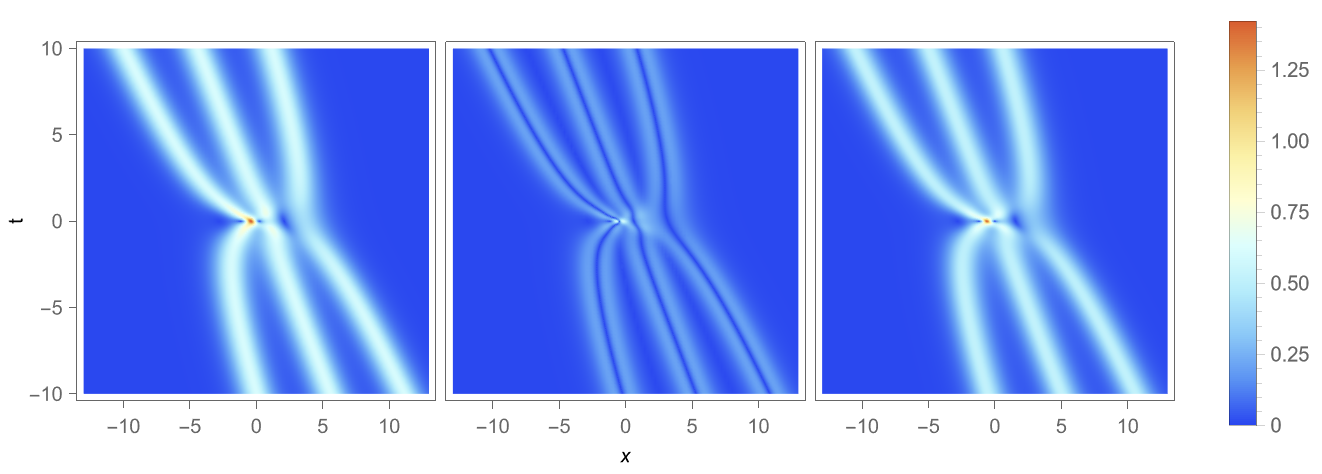}
\end{subfigure}
\caption{ \label{D31}Density structures of three components $|q_1|$, $|q_0|$ and $ |q_{-1}|$ (from left to right) as $N=1$, $\lambda_1=\frac{1}{8}+\frac{\mathrm{i}}{4}$, $m_1=2$, $\mathbf{f}(\lambda)=\begin{pmatrix}
                                             1 & 1 \\
                                             1 & 1
                                           \end{pmatrix}$ (\textbf{top}),  $\mathbf{f}(\lambda)=\begin{pmatrix}
                                             3 & 8 \\
                                             8 & 4
                                           \end{pmatrix}$ (\textbf{middle}), $\mathbf{f}(\lambda)=\begin{pmatrix}
                                             8 & 2 \\
                                             2 & 4
                                           \end{pmatrix}$ (\textbf{bottom}).  }
\end{figure}

\begin{figure}[ht]
\centering
\begin{subfigure}{\textwidth}
\centering
\includegraphics[scale=.5]{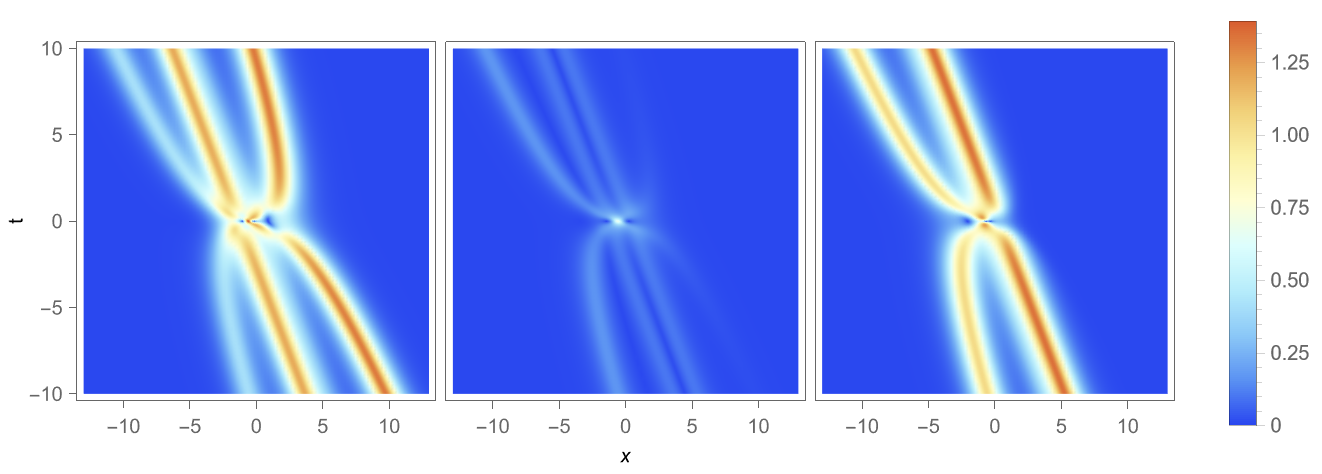}
\end{subfigure}
\begin{subfigure}{\textwidth}
\centering
\includegraphics[scale=.5]{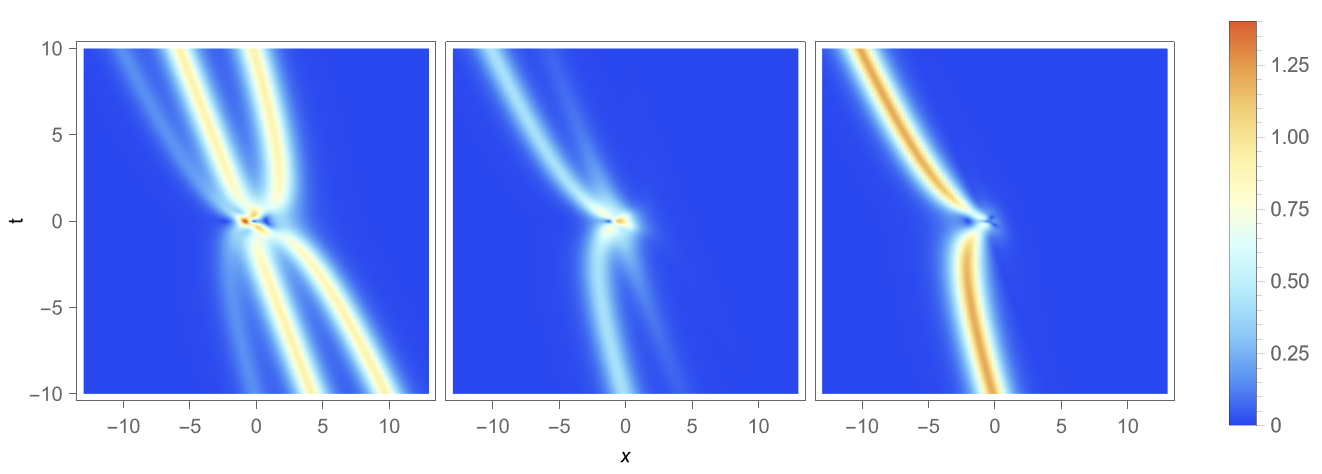}
\end{subfigure}
\begin{subfigure}{\textwidth}
\centering
\includegraphics[scale=.5]{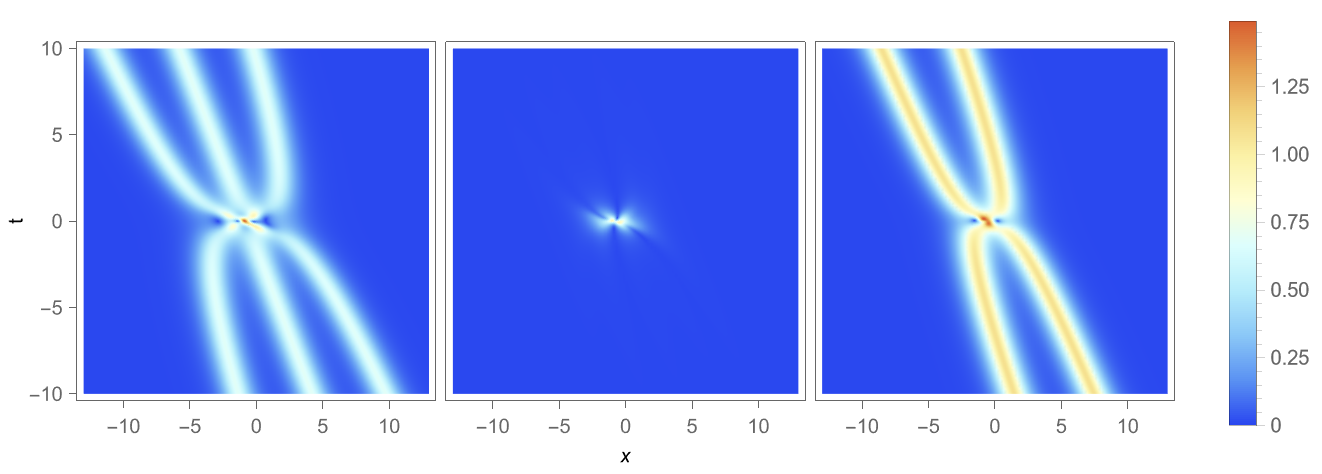}
\end{subfigure}
\begin{subfigure}{\textwidth}
\centering
\includegraphics[scale=.5]{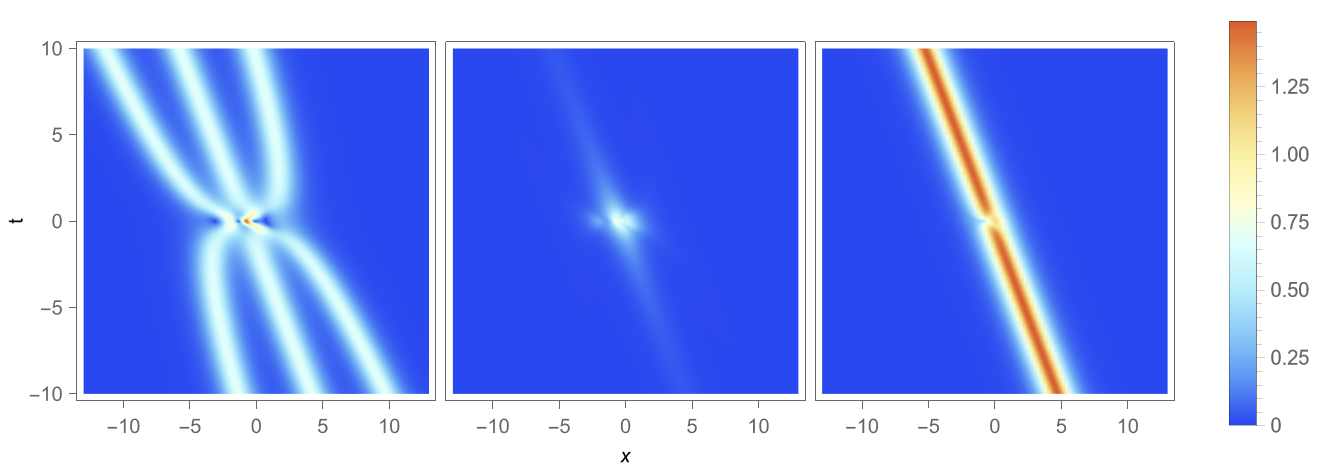}
\end{subfigure}
\caption{ \label{D32}Density structures of three components $|q_1|$, $|q_0|$ and $ |q_{-1}|$ (from left to right) as $N=1$, $\lambda_1=\frac{1}{8}+\frac{\mathrm{i}}{4}$, $m_1=2$, $\mathbf{f}(\lambda)=\begin{pmatrix}
                                            1 & \lambda-\lambda_1 \\
                                            \lambda-\lambda_1 & 0
                                          \end{pmatrix}$ (\textbf{1st row}),   $\mathbf{f}(\lambda)=\begin{pmatrix}
                                            1 & (\lambda-\lambda_1)^2 \\
                                            (\lambda-\lambda_1)^2 & 0
                                          \end{pmatrix}$ (\textbf{2nd row}), $\mathbf{f}(\lambda)=\begin{pmatrix}
                                            1 & (\lambda-\lambda_1)^2 \\
                                            (\lambda-\lambda_1)^2 & \lambda-\lambda_1
                                          \end{pmatrix}$ (\textbf{3rd row}), $\mathbf{f}(\lambda)=\begin{pmatrix}
                                            1 & (\lambda-\lambda_1)^2 \\
                                            (\lambda-\lambda_1)^2 & (\lambda-\lambda_1)^2
                                          \end{pmatrix}$ (\textbf{4th row}). }
\end{figure}

\begin{figure}[ht]
\centering
\begin{subfigure}{\textwidth}
\centering
\includegraphics[scale=.5]{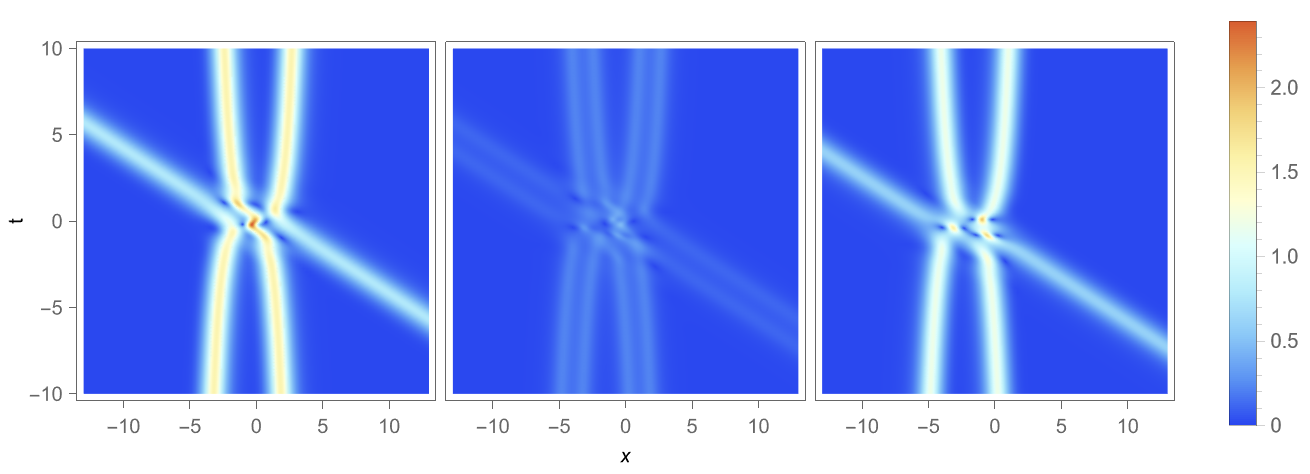}
\end{subfigure}
\begin{subfigure}{\textwidth}
\centering
\includegraphics[scale=.5]{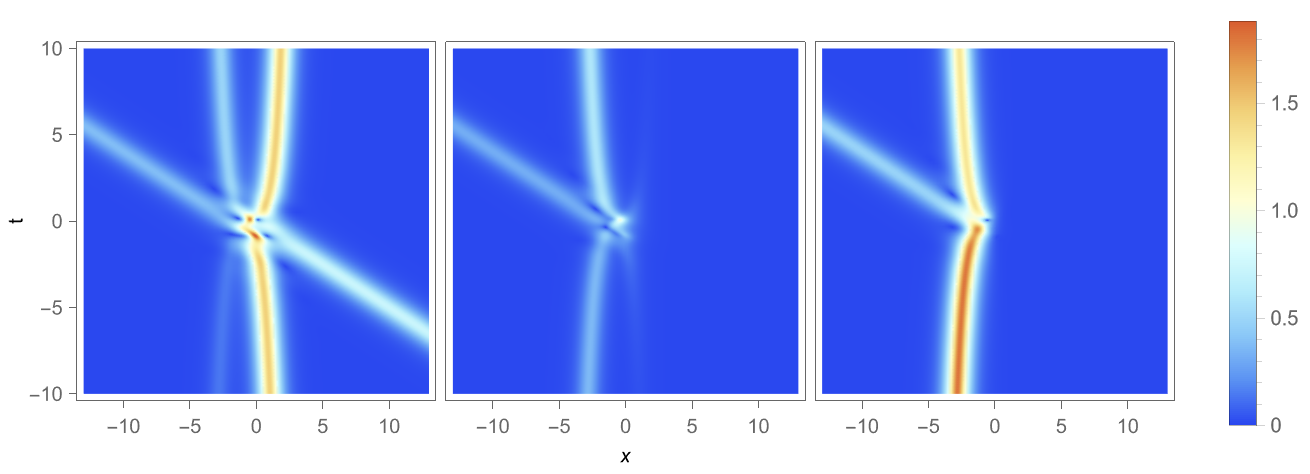}
\end{subfigure}
\begin{subfigure}{\textwidth}
\centering
\includegraphics[scale=.5]{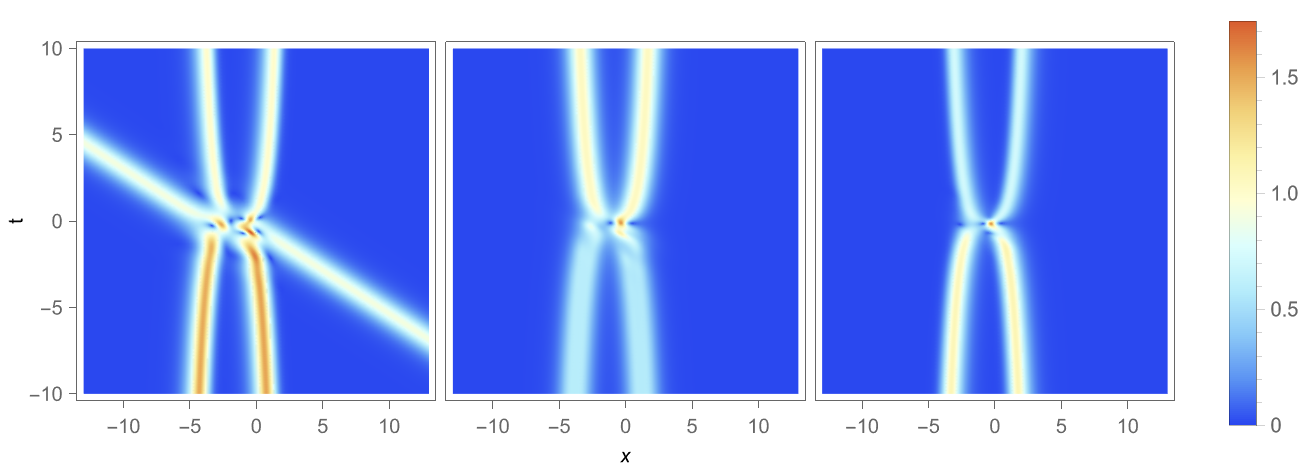}
\end{subfigure}
\begin{subfigure}{\textwidth}
\centering
\includegraphics[scale=.5]{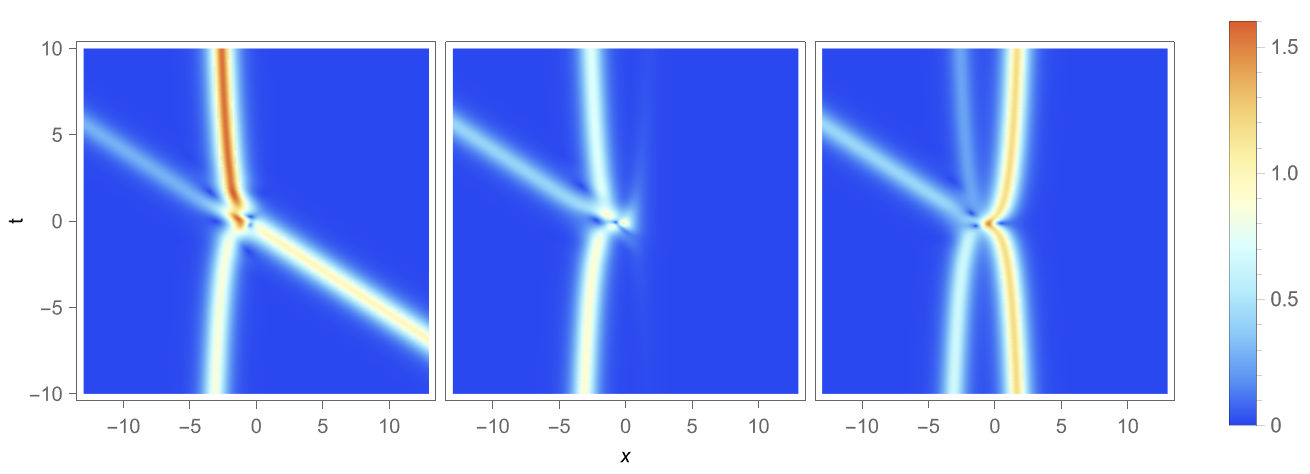}
\end{subfigure}
\caption{ \label{D221}Density structures of three components $|q_1|$, $|q_0|$ and $ |q_{-1}|$ (from left to right) as $N=2$, $\lambda_1=\mathrm{i}$, $\lambda_2=\frac{1+\mathrm{i}}{2}$, $m_1=1$, $m_2=0$, $\mathbf{f}(\lambda)=\begin{pmatrix}
                                            5 & 1 \\
                                            1 &0
                                          \end{pmatrix}$ (\textbf{1st row}),   $\mathbf{f}(\lambda)=\begin{pmatrix}
                                            1 & (\lambda-\lambda_1)(\lambda-\lambda_2) \\
                                          (\lambda-\lambda_1)(\lambda-\lambda_2)& 0
                                          \end{pmatrix}$ (\textbf{2nd row}), $\mathbf{f}(\lambda)=\begin{pmatrix}
                                            (\lambda-\lambda_1)^2  & \lambda-\lambda_2\\
                                             \lambda-\lambda_2 &  \lambda-\lambda_2
                                          \end{pmatrix}$ (\textbf{3rd row}), $\mathbf{f}(\lambda)=\begin{pmatrix}
                                            (\lambda-\lambda_1)^2   & (\lambda-\lambda_1)(\lambda-\lambda_2)\\
                                           (\lambda-\lambda_1)(\lambda-\lambda_2) & \lambda-\lambda_2
                                          \end{pmatrix}$ (\textbf{4th row}). }
\end{figure}

\begin{figure}[ht]
\centering
\begin{subfigure}{\textwidth}
\centering
\includegraphics[scale=.5]{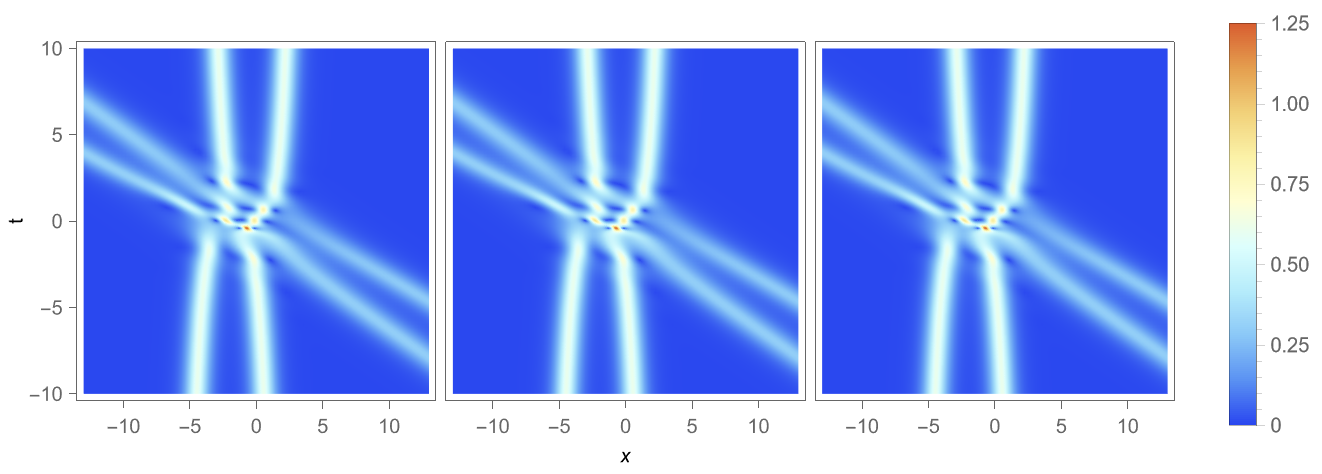}
\end{subfigure}
\begin{subfigure}{\textwidth}
\centering
\includegraphics[scale=.5]{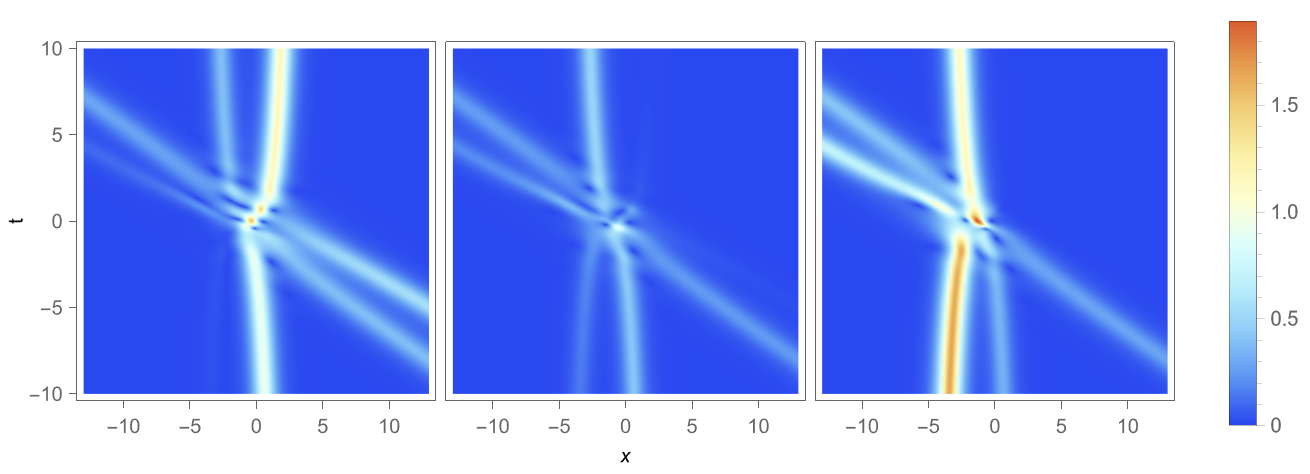}
\end{subfigure}
\begin{subfigure}{\textwidth}
\centering
\includegraphics[scale=.5]{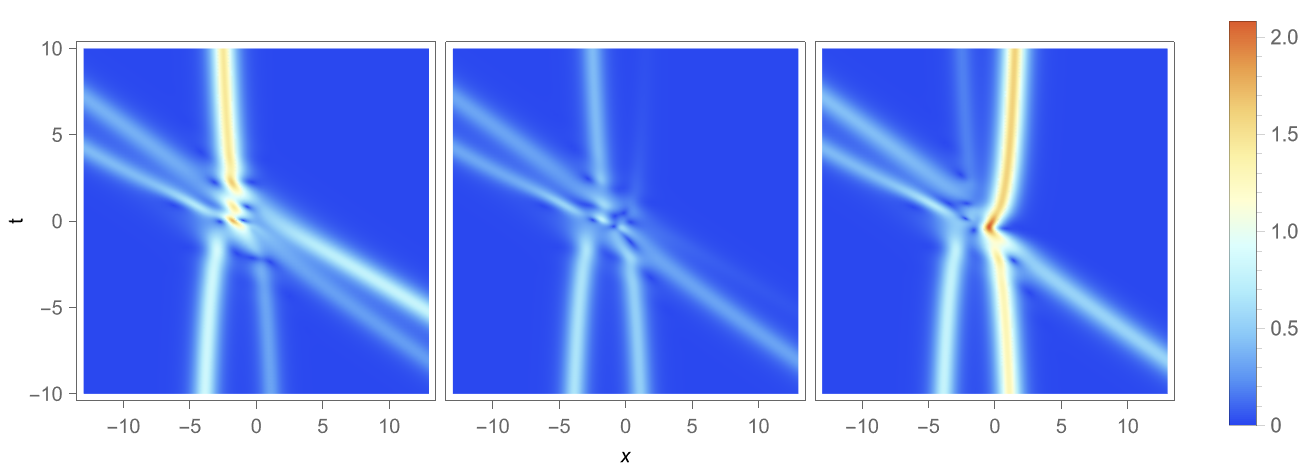}
\end{subfigure}
\begin{subfigure}{\textwidth}
\centering
\includegraphics[scale=.5]{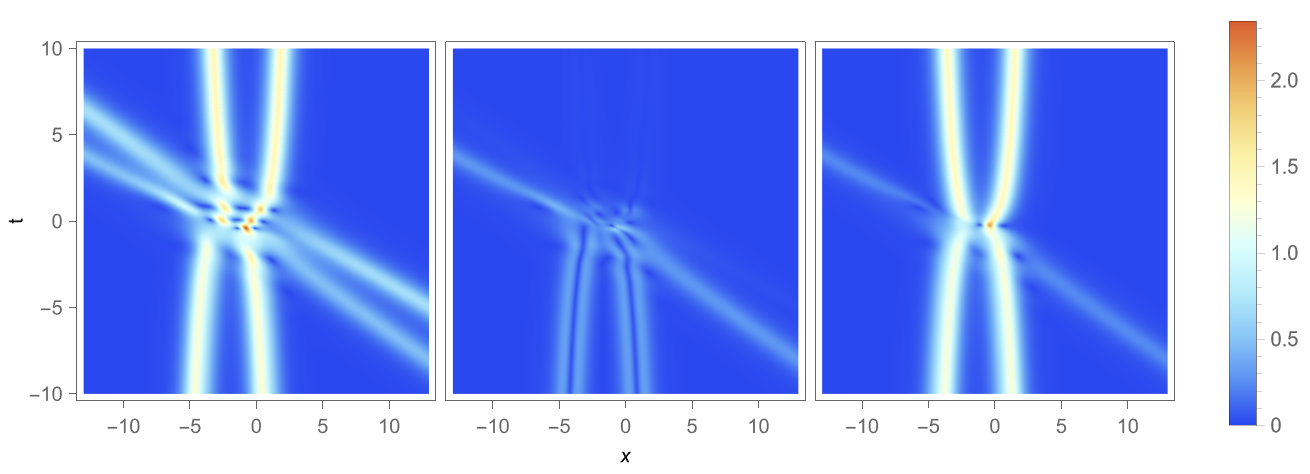}
\end{subfigure}
\caption{ \label{D222}Density structures of three components $|q_1|$, $|q_0|$ and $ |q_{-1}|$ (from left to right) as $N=2$, $\lambda_1=\mathrm{i}$, $\lambda_2=\frac{1+\mathrm{i}}{2}$, $m_1=1$, $m_2=1$,  $\mathbf{f}(\lambda)=\begin{pmatrix}
                                            1 & 1 \\
                                            1 &1
                                          \end{pmatrix}$ (\textbf{1st row}),   $\mathbf{f}(\lambda)=\begin{pmatrix}
                                            1 & (\lambda-\lambda_1)(\lambda-\lambda_2) \\
                                           (\lambda-\lambda_1)(\lambda-\lambda_2) & 0
                                          \end{pmatrix}$ (\textbf{2nd row}), $\mathbf{f}(\lambda)=\begin{pmatrix}
                                            (\lambda-\lambda_1)^2  & (\lambda-\lambda_1)(\lambda-\lambda_2) \\
                                            (\lambda-\lambda_1)(\lambda-\lambda_2)  & (\lambda-\lambda_2)^2
                                          \end{pmatrix}$ (\textbf{3rd row}), $\mathbf{f}(\lambda)=\begin{pmatrix}
                                            1 & (\lambda-\lambda_1)(\lambda-\lambda_2) \\
                                            (\lambda-\lambda_1)(\lambda-\lambda_2)  & (\lambda-\lambda_2)^2
                                          \end{pmatrix}$ (\textbf{4th row}). }
\end{figure}

In the following, we explore  different possibilities for  multiple higher-order poles solutions. For $N=1$, we derive several third-order bounded state soliton solutions and display the density structures of $(|q_1|,|q_0|,|q_{-1}|)$ by choosing $\lambda_1=\frac{1}{8}+\frac{\mathrm{i}}{4}$ with multiplicity 3, i.e., $m_1=2$, and various $\mathbf{f}(\lambda)$, as shown in Figs\,\ref{D31}-\ref{D32}. All of the solutions displayed in Fig.\,\ref{D31} are  non-degenerate bounded state solitons. In particular, $q_{\pm1}$ in the middle case and $q_0$ in the bottom case are double-humped solitons, while the others are   singe-humped soltions. Some components  displayed in Fig.\,\ref{D32} are third-order degenerate bounded state solitons, such as  one-soliton solution, second-order bounded state soliton solutions and localized solutions.  For $N=2$, we set $\lambda_1=\mathrm{i}$, $\lambda_2=\frac{1+\mathrm{i}}{2}$, and display the collisions of several second-order bounded state solitons and one-solitons by choosing  $m_1=1$, $m_2=0$, and various $\mathbf{f}(\lambda)$, as shown in Fig\,\ref{D221}. Moreover, the collisions of two second-order solitons are presented in Fig.\,\ref{D222} by choosing $m_1=1$, $m_2=1$, and various $\mathbf{f}(\lambda)$.

   By choosing suitable values for  $\lambda_1,\ldots,\lambda_N$, $m_1,\ldots, m_N$ and $\mathbf{f}(\lambda)$, one can obtain a wide range of structures for the triplet $(|q_1|,|q_0|,|q_{-1}|)$. Based on this, it is conjectured that all vanishing-at-infinity soliton solutions of the spin-1 GP equation can be constructed by selecting an appropriate combination of  $\{\lambda_1,\ldots,\lambda_N,m_1,\ldots,m_N,\mathbf{f}(\lambda)\}$.
\section*{Data Availability Statements}
All data  analysed during this study are included in this published article.
\section*{Conflict of interests}
On behalf of all authors, the corresponding author states that there is no conflict of interest.

\section*{Acknowledgment}
This work was supported by National Natural Science Foundation of China (Grant Nos. 12171439, 12101190, 11931017).
\newpage
\bibliographystyle{siam}

\bibliography{references}
\end{document}